\documentclass[aps,pre,twocolumn,superscriptaddress,longbibliography]{revtex4-2}
\usepackage{amsmath}
\usepackage{amsfonts}
\usepackage{amssymb}
\usepackage{lmodern,dsfont}
\usepackage{graphicx}
\usepackage[usenames,dvipsnames]{xcolor}
\usepackage{bm}
\usepackage[english=american]{csquotes}
\usepackage{xr}

\newcommand{\Av}[1]{\left\langle #1 \right\rangle}
\newcommand{\av}[1]{\langle #1 \rangle}
\newcommand{\n}{\nonumber}
\newcommand{\nn}{\nonumber \\}

\newcommand{\grad}{\bm{\nabla}}

\renewcommand{\eqref}[1]{Eq.~(\ref{#1})}

\begin{document}
\author{Andreas Dechant}
\affiliation{Department of Physics \#1, Graduate School of Science, Kyoto University, Kyoto 606-8502, Japan}

\title{Upper bounds on entropy production in diffusive dynamics}

\begin{abstract}
Based on a variational expression for the steady-state entropy production rate in overdamped Langevin dynamics, we derive concrete upper bounds on the entropy production rate in various physical settings.
For particles in a thermal environment and driven by non-conservative forces, we show that the entropy production rate can be upper bounded by considering only the statistics of the driven particles.
We use this finding to argue that the presence of non-driven, passive degrees of freedom generally leads to decreased dissipation.
Another upper bound can be obtained only in terms of the variance of the non-conservative force, which leads to a universal upper bound for particles that are driven by a constant force that is applied in a certain region of space.
Extending our results to systems attached to multiple heat baths or with spatially varying temperature and/or mobility, we show that the temperature difference between the heat baths or the gradient of the temperature can be used to upper bound the entropy production rate.
We show that most of these results extend in a straightforward way to underdamped Langevin dynamics and demonstrate them in three concrete examples.
\end{abstract}

\maketitle

\section{Introduction} \label{sec-intro}

From a thermodynamic point of view, the distinction between equilibrium and out-of-equilibrium systems is the irreversibility of the latter.
This irreversibility is quantified by entropy production, which measures the inevitable increase in the entropy of the universe during any out-of-equilibrium process.
In practice, entropy production leads to dissipation, that is, a loss of energy into the environment of the system, and therefore, one frequently seeks to minimize the entropy production involved in a given process.
Conversely, far-from-equilibrium systems can exhibit a rich variety of non-trivial behaviors, which are not possible in or near equilibrium, such as non-monotonic or even negative transport coefficients \cite{Eic02,Ros05}, persistent oscillations \cite{Obe22} or spontaneous emergence of patterns \cite{Cro93,Gol99}.

A recent trend in non-equilibrium statistical physics is to investigate lower bounds on the entropy production in terms of observable quantities, which can be used to estimate entropy production from measured data \cite{Bar15,Dec17,Pie17,Li19,Sei19,Hor20,Ski21,Dec21c,Mee22,Har22}.
From a more fundamental perspective, such lower bounds also quantify how much entropy production is strictly necessary to observe a given non-equilibrium phenomenon.
Given the recent success of this approach, it seems natural to ask the opposite question: Given our knowledge about a particular system, can we derive an \emph{upper} bound on the entropy production?
Clearly, this question cannot be answered from measured data alone: If there is some non-equilibrium process occurring in the system that is not reflected in the measurement, then we have no way of estimating its contribution to the entropy production.

In many cases, a physical system is driven out of equilibrium by an externally controlled operation, such as applying a bias force or a temperature gradient.
In such cases, we know the thermodynamic forces driving the system out of equilibrium. 
We show that this additional information about the driving forces can be used, potentially together with measured data, to obtain upper bounds on the entropy production.
In particular, for Brownian particles in contact with a thermal environment (characterized by temperature $T$ and friction coefficient $\gamma$), that are driven into a non-equilibrium steady state by a non-conservative force-field $\bm{F}^\text{nc}$, we obtain a simple upper bound on the rate of entropy production,
\begin{align}
\sigma_\text{st} \leq \frac{1}{\gamma T} \Av{ \big\Vert \bm{F}^\text{nc} \big\Vert^2} ,
\end{align}
where $\av{\ldots}$ denotes an ensemble average and $\Vert \ldots \Vert$ is the vector norm.
This bound, which holds for both over- and underdamped motion, places an upper limit on the dissipation incurred due the non-conservative force, which only depends on the average magnitude of the driving force and the properties of the thermal environment.
As we will discuss in the following, the bound can be tightened using additional information and can be extended to cases where the thermodynamic force arises due to several heat baths or a temperature gradient.

The structure of the paper is as follows:
In Section \ref{sec-setup} we introduce the basic setup of overdamped Langevin dynamics and discuss a variational formula for the entropy production rate.
We apply this variational formula to systems of particles driven by non-conservative forces in Section \ref{sec-thermal} to obtain various upper bounds on the entropy production rate for specific situations.
In Sections \ref{sec-temp-gradient} and \ref{sec-temp-gradient} we extend the bounds to systems in contact with several heat baths or in the presence of temperature gradients, respectively.
Section \ref{sec-periodic} discusses the necessary modifications of the bounds when dealing with system with periodic boundary conditions.
While most of the derivations are done with overdamped Langevin dynamics in mind, we show in Section \ref{sec-underdamped} that almost all derived upper bounds also apply to underdamped systems with no or little modification.
Finally, in Section \ref{sec-demonstration}, we compare the upper bounds to the exact value of the entropy production for interacting driven Brownian particles, a temperature ratchet, and underdamped motion in periodic potentials.

\section{Setup and variational principle} \label{sec-setup}

We consider a system of $N$ overdamped degrees of freedom, whose motion is described by the overdamped Langevin equations \cite{Cof17}
\begin{align}
\dot{\bm{x}}(t) = \bm{a}(\bm{x}(t)) + \bm{G}(\bm{x}(t)) \cdot \bm{\xi}(t ) \label{langevin}.
\end{align}
Here $\bm{a}(\bm{x})$ is a drift vector, $\bm{G}(\bm{x})$ is a rank $N$ coupling matrix describing the coupling to the $N$ Gaussian white noises $\bm{\xi}(t)$ and $\cdot$ denotes the Ito-product.
We assume that the drift vector and coupling matrix are such that the system reaches a steady state at long time.
The steady-state probability density $p_\text{st}(\bm{x})$ is determined from the Fokker-Planck equation \cite{Ris86}
\begin{align}
0 &= - \grad \big( \bm{\nu}_\text{st}(\bm{x}) p_\text{st}(\bm{x}) \big) \qquad \text{with} \label{fpe-steady} \\
\bm{\nu}_\text{st}(\bm{x}) &= \bm{a}(\bm{x}) - \grad \bm{B}(\bm{x}) - \bm{B}(\bm{x}) \grad \ln p_\text{st}(\bm{x}) \n .
\end{align}
Here $\bm{B}(\bm{x}) = 2 \bm{G}(\bm{x}) \bm{G}(\bm{x})^\text{T}$ is the positive definite diffusion matrix and $\grad \bm{B}(\bm{x})$ denotes the vector with entries $\partial_{x_j} B_{ij}(\bm{x})$, where the sum over repeated indices is implied.
The quantity $\bm{\nu}_\text{st}(\bm{x})$ is called the local mean velocity, it characterizes the flows in the steady state of the system, which is thus generally a non-equilibrium steady state with entropy production rate \cite{Sek10,Sei12}
\begin{align}
\sigma_\text{st} = \Av{\bm{\nu}_\text{st} \bm{B}^{-1} \bm{\nu}_\text{st}} \label{entropy},
\end{align}
where $\av{\ldots}$ denotes the average with respect to the steady-state probability density.
In the following, we will focus on systems under natural or reflecting boundary conditions, such that the probability current $\bm{j}_\text{st}(\bm{x}) = \bm{\nu}_\text{st}(\bm{x}) p_\text{st}(\bm{x})$ vanishes at the boundaries or as $\Vert \bm{x} \Vert \rightarrow \infty$.
We will remark on systems with periodic boundary conditions later.
Due to the steady state condition \eqref{fpe-steady}, we have for arbitrary gradient fields $\grad \psi(\bm{x})$,
\begin{align}
\Av{\grad \psi \bm{\nu}_\text{st}} &= \int d\bm{x} \ p_\text{st}(\bm{x}) \bm{\nu}_\text{st}(\bm{x}) \grad \psi(\bm{x}) \label{orthogonality} \\
& = - \int d\bm{x} \ \psi(\bm{x}) \grad \big( \bm{\nu}_\text{st}(\bm{x}) p_\text{st}(\bm{x}) \big) = 0 \n .
\end{align}
That is, the local mean velocity is orthogonal to all gradient fields with respect to the inner product over the space of vector fields $\Av{\bm{u},\bm{v}} = \Av{\bm{u} \bm{v}}$.
Next, let us consider the functional
\begin{align}
\Psi[\psi] = \Av{ \big( \bm{\nu}_\text{st} + \bm{B} \grad \psi \big) \bm{B}^{-1} \big( \bm{\nu}_\text{st} + \bm{B} \grad \psi \big)} .
\end{align}
Using \eqref{orthogonality}, we can write this as
\begin{align}
\Psi[\psi] = \sigma_\text{st} + \Av{\grad \psi \bm{B} \grad \psi} .
\end{align}
Since the second term is positive and vanishes for any constant function $\grad \psi(\bm{x}) = 0$, we have at the identity,
\begin{align}
\sigma_\text{st} = \inf_{\psi} \Psi[\psi] \label{entropy-variational} .
\end{align}
While at this level, the variational principle is almost trivial, we will see in the following that it is useful for deriving upper bounds on the entropy production rate in terms of computable quantities.
At this point, we note that \eqref{entropy-variational} gives an upper bound on the entropy production rate for any choice of $\psi(\bm{x})$,
\begin{align}
\sigma_\text{st} \leq \Av{ \big( \bm{\nu}_\text{st} + \bm{B} \grad \psi(\bm{x}) \big) \bm{B}^{-1} \big( \bm{\nu}_\text{st} + \bm{B} \grad \psi(\bm{x}) \big)} .
\end{align}
We remark that a similar variational formula has been derived for the housekeeping entropy in the Maes-Neto{\v{c}}n{\`y} decomposition \cite{Mae14}.
In Ref.~\cite{Dec22}, it was shown that we have
\begin{align}
\sigma_\text{hk} = \inf_{\psi} \Av{\big( \bm{\nu}_t + \bm{B}_t \grad \psi \big) \bm{B}_t^{-1} \big( \bm{\nu}_t + \bm{B}_t \grad \psi(\bm{x}) \big)}_t \label{entropy-variational-time} .
\end{align}
Here $\bm{\nu}_t(\bm{x})$ is the time-dependent local mean velocity,
\begin{align}
\partial_t p_t(\bm{x}) &= - \grad \big( \bm{\nu}_t(\bm{x}) p_t(\bm{x}) \big) \qquad \text{with} \label{fpe} \\
\bm{\nu}_t(\bm{x}) &= \bm{a}_t(\bm{x}) - \grad \bm{B}_t(\bm{x}) - \bm{B}_t(\bm{x}) \grad \ln p_t(\bm{x}) \n ,
\end{align}
and the average is taken with respect to the time-dependent probability density $p_t(\bm{x})$.
The similarity between \eqref{entropy-variational-time} and \eqref{entropy-variational} implies that all results that are derived from \eqref{entropy-variational} for the steady state entropy in the following also apply without change to the housekeeping entropy in arbitrary time-dependent dynamics.

\section{Thermal systems} \label{sec-thermal}

\subsection{General non-conservative forces}
An important class of physical systems described by \eqref{langevin} are Brownian particles in contact with an equilibrium environment with temperature $T$.
The interactions between the particles and the effect of external forces result in a total force $\bm{F}(\bm{x})$, while the interaction between the particles and the environment is described by a positive definite mobility matrix $\bm{\mu}$.
The equations of motion are then
\begin{align}
\dot{\bm{x}} = \bm{\mu} \bm{F}(\bm{x}) + \sqrt{2 \bm{\mu} T} \bm{\xi}(t) .
\end{align}
The matrix $\sqrt{\bm{\mu}}$ is defined as the unique positive definite matrix with $\sqrt{\bm{\mu}}\sqrt{\bm{\mu}} = \bm{\mu}$.
In this case, the diffusion matrix is $\bm{\mu} T$ and the steady-state local mean velocity is given by
\begin{align}
\bm{\nu}_\text{st}(\bm{x}) &= \bm{\mu} \big( \bm{F}(\bm{x}) - T \grad \ln p_\text{st}(\bm{x}) \big) .
\end{align}
If all forces in the system are conservative, we can write $\bm{F}(\bm{x}) = -\grad U(\bm{x})$ and the steady-state is given by the Boltzmann-Gibbs equilibrium,
\begin{align}
p_\text{st}(\bm{x}) = Z^{-1} e^{-\frac{U(\bm{x})}{T}} \quad \text{with} \quad Z = \int d\bm{x} \ e^{-\frac{U(\bm{x})}{T}} ,
\end{align}
and the local mean velocity and entropy production rate vanish.
Thus, for systems in contact with a thermal environment, conservative forces are equivalent to relaxation to thermal equilibrium.
In the presence of non-conservative forces that cannot be written as the gradient of a potential $U(\bm{x})$, the system will relax to a non-equilibrium steady state with a non-zero rate of entropy production, which in this case is equal to the rate of heat dissipation into the environment,
\begin{align}
\sigma_\text{st} = \frac{\dot{Q}_\text{diss}}{T} .
\end{align}
Let us now apply \eqref{entropy-variational} to this situation.
We have
\begin{align}
\sigma_\text{st} &= \frac{1}{T} \inf_{\psi} \Big\langle \big( \bm{F} - T \grad \ln p_\text{st} + T \grad \psi \big) \\
& \hspace{2cm} \times \bm{\mu} \big( \bm{F} - T \grad \ln p_\text{st} + T \grad \psi \big) \Big\rangle . \n
\end{align}
Since the infimum is taken over all gradient fields, we can absorb the terms involving the steady-state probability density into $\psi(\bm{x})$,
\begin{align}
\sigma_\text{st} &= \frac{1}{T} \inf_{\psi} \Big\langle \big( \bm{F} + T \grad \psi \big) \bm{\mu} \big( \bm{F} + T \grad \psi \big) \Big\rangle . 
\end{align}
Next, we split the forces into a conservative and non-conservative part as $\bm{F}(\bm{x}) = -\grad U(\bm{x}) + \bm{F}^\text{nc}(\bm{x})$, again absorbing the gradient term into $\psi(\bm{x})$,
\begin{align}
\sigma_\text{st} &= \frac{1}{T} \inf_{\psi} \Big\langle \big( \bm{F}^\text{nc} + T \grad \psi \big) \bm{\mu} \big( \bm{F}^\text{nc} + T \grad \psi \big) \Big\rangle . 
\end{align}
Defining $V(\bm{x}) = T \psi(\bm{x})$, we then arrive at
\begin{align}
\sigma_\text{st} &= \frac{1}{T} \inf_{V} \Big\langle \big( \bm{F}^\text{nc} + \grad V \big) \bm{\mu} \big( \bm{F}^\text{nc} + \grad V \big) \Big\rangle \label{entropy-variational-nonconservative} . 
\end{align}
This corresponds to a minimization over conservative forces, that is, finding the conservative force $-\grad V(\bm{x})$ that is most similar to the non-conservative force $\bm{F}^\text{nc}(\bm{x})$.
Note that the expression \eqref{entropy-variational-nonconservative} does not explicitly involve the potential forces, which only appear via their influence on the steady-state probability with respect to which the average is taken.
In many physical situations, the non-conservative force corresponds to an externally applied force, and therefore, its functional form is often known.
By contrast, the conservative force generally also include interactions between the particles, as well as interactions with e.~g.~obstacles in the environment, whose precise form is often not known.
\eqref{entropy-variational-nonconservative} allows us to calculate the entropy production rate from the knowledge of only the non-conservative forces by computing ensemble averages.
Rescaling $V(x) \rightarrow \alpha V(x)$ and minimizing with respect to $\alpha$, we find the equivalent expression,
\begin{align}
\sigma_\text{st} &= \frac{1}{T} \inf_{V} \bigg( \Av{\bm{F}^\text{nc}\bm{\mu} \bm{F}^\text{nc}} - \frac{\Av{\bm{F}^\text{nc}\bm{\mu} \grad V}^2}{\Av{\grad V \bm{\mu} \grad V }} \bigg) \label{entropy-variational-nonconservative-2} , 
\end{align}
which generally provides a tighter bound than \eqref{entropy-variational-nonconservative} for an arbitrary choice of $V(\bm{x})$, since we already optimized with respect to the overall magnitude of $V(\bm{x})$.
Choosing $V(\bm{x}) = 0$ in \eqref{entropy-variational-nonconservative}, we obtain our first upper bound on the entropy production rate,
\begin{align}
\sigma_\text{st} \leq \frac{1}{T} \Av{\bm{F}^\text{nc} \bm{\mu} \bm{F}^\text{nc}} \label{entropy-bound-nonconservative-simple} .
\end{align}
The right-hand side corresponds to the entropy production that would be obtained if the entire non-conservative force was converted into a current, $\bm{\nu}_\text{st}(\bm{x}) = \bm{\mu} \bm{F}^\text{nc}(\bm{x})$.
This bound ascertains that the entropy production rate cannot exceed the overall magnitude of the non-conservative force.
A tighter bound, which likewise is expressed only in terms of the non-conservative force can be obtained from \eqref{entropy-variational-nonconservative} by choosing $V = -\bm{x} \Av{\bm{F}^\text{nc}}_\text{st}$,
\begin{align}
\sigma_\text{st} \leq \frac{1}{T} \Big\langle \big( \bm{F}^\text{nc} - \Av{\bm{F}^\text{nc}} \big) \bm{\mu} \big( \bm{F}^\text{nc} - \Av{\bm{F}^\text{nc}} \big) \Big\rangle \label{entropy-bound-nonconservative} .
\end{align}
This implies that the entropy production is upper bounded by the deviation of the non-conservative force from its average value.
This bound takes an even simpler shape if the non-conservative force is a constant force $\bm{F}^\text{nc}(\bm{x}) = \bm{F}_0 \chi_\Omega(\bm{x})$ that is acting over some finite region $\Omega$ of the configuration space, where $\chi_\Omega(\bm{x}) = 1$ if $\bm{x} \in \Omega$ and $\chi_\Omega(\bm{x}) = 0$ otherwise.
In this case, we have
\begin{align}
\sigma_\text{st} \leq \frac{\bm{F}_0 \bm{\mu} \bm{F}_0}{T} P(\Omega) \big(1-P(\Omega) \big) \label{entropy-bound-constant},
\end{align}
where $P(\Omega)$ denotes the probability of finding the system in $\Omega$.
Obviously, this expression vanishes if $P(\Omega) = 0$, when the system is never driven by the non-conservative force, or if $P(\Omega) = 1$, since a constant force acting over the entire configuration space is conservative $\bm{F}_0 = - \grad (\bm{x} \bm{F}_0)$.
The expression on the right-hand side is maximal if $P(\Omega) = 1/2$ and we obtain the global upper bound
\begin{align}
\sigma_\text{st} \leq \frac{\bm{F}_0 \bm{\mu} \bm{F_0}}{4 T}.
\end{align}
This gives an upper bound on the entropy production of a system driven by a constant magnitude non-conservative force, which depends only on the magnitude of the force, the mobility and the temperature.

We can also compute the entropy production rate from the work done by the non-conservative force \cite{Sek10,Sei12},
\begin{align}
\mathcal{W} = \int_0^\tau dt \ \bm{F}^\text{nc}(\bm{x}(t)) \circ \dot{\bm{x}}(t) \label{work} .
\end{align}
Here, $\dot{\bm{x}}(t)$ is the velocity given by \eqref{langevin} and $\circ$ denotes the Stratonovich product.
In the steady-state, the average work is related to the entropy production rate by
\begin{align}
\sigma_\text{st} = \frac{\Av{\mathcal{W}}}{T \tau} \label{entropy-work} .
\end{align}
Just like \eqref{entropy-variational-nonconservative}, this can be used to calculate the entropy production rate from the knowledge of the non-conservative force.
At first glance, \eqref{work} appears to be simpler, since it does not involve any variational expression.
However, it requires the velocity $\dot{\bm{x}}(t)$, which can only be measured accurately if the time-resolution is sufficient to resolve all the timescales in the dynamics, which may be challenging in particular in strongly interacting systems.
By contrast, \eqref{entropy-variational-nonconservative} relies on sampling statistics from the steady-state probability density, which does not place any restrictions on the time-resolution.
Therefore, \eqref{entropy-variational-nonconservative} may yield a more accurate result in situations where the velocity cannot be resolved well, at the cost that we have to perform a minimization with respect to $V(\bm{x})$.
Expanding $V(\bm{x})$ into a set of $Q$ basis functions, $V(\bm{x}) = \sum_{k=1}^Q c_q \eta_q(\bm{x})$, \eqref{entropy-variational-nonconservative} requires computing the $(Q+2)(Q+1)/2$ coefficients
\begin{gather}
C_{00} = \Av{\bm{F}^\text{nc} \bm{\mu} \bm{F}^\text{nc}}, \quad C_{0q} = \Av{\bm{F}^\text{nc} \bm{\mu} \grad \eta_q},\\
 C_{q r} = \Av{\grad \eta_q \bm{\mu} \grad \eta_r} , \n
\end{gather}
which can be done by evaluating the functions for the measured values of $\bm{x}(t)$ and averaging.
In terms of the coefficient matrix $\bm{C}$, the minimizer of \eqref{entropy-variational-nonconservative}, and thus the resulting estimate on the entropy production rate, is given by
\begin{align}
\hat{\sigma}_\text{st} = C_{00} - \sum_{q,r = 1}^K C_{0q} \big(\bm{C}^{-1} \big)_{qr} C_{0r} \label{entropy-basis} .
\end{align}
For an finite set of basis functions (and ignoring other sources of errors), this is an upper estimate on the entropy production rate, $\sigma_\text{st} \leq \hat{\sigma}_\text{st}$, which will converge to the true value as the number of basis functions is increased, provided that they form a complete basis of the configuration space.

\subsection{Driven and passive particles}
We next consider a system of $M+K$ Brownian particles in $n$ dimensions in contact with a heat bath at temperature $T$, whose positions at time $t$ we label by $\bm{x}(t) = (\bm{y}(t),\bm{z}(t)) = (\bm{y}_1(t),\ldots,\bm{y}_M(t),\bm{z}_1(t),\ldots,\bm{z}_K(t))$ with $\bm{y}_1(t) = (y_{1,1}(t),\ldots,y_{1,n}(t))$, that is, $N = (M+K)n$ degrees of freedom.
Conservative external forces and interactions between the particles are encoded in the potential $U(\bm{x}) = U(\bm{y},\bm{z})$, which may depend on both sets of particles.
In addition, there is a nonconservative force $\bm{F}_\text{d}^\text{nc}(\bm{y})$ that acts on $M$ of the particles, which we refer to as the driven particles, while the remaining $K$ particles, which are only subject to potential forces, are called passive.
We further assume that the mobility matrix is block-diagonal, with its first $M \times M$ block equal to the mobility matrix $\bm{\mu}_\text{d}$ of the driven particles, and the remaining $K \times K$ block equal to the mobility matrix $\bm{\mu}_\text{p}$ of the passive particles.
This means that the environment does not induce any interactions between the driven and non-driven particles.
In \eqref{entropy-variational-nonconservative}, we now choose $V(\bm{x}) = V_\text{d}(\bm{y})$, that is, we allow the potential to only depend on the driven degrees of freedom.
Restricting the domain of the minimization in that way results in an upper bound,
\begin{align}
\sigma_\text{st} \leq \frac{1}{T} \inf_{V_\text{d}} \Big\langle \big( \bm{F}_\text{d}^\text{nc} + \grad_y V_\text{d} \big) \bm{\mu} \big( \bm{F}^\text{nc} + \grad_y V_\text{d} \big) \Big\rangle .
\end{align}
Since the mobility matrix is block-diagonal and the vector $\bm{F}_\text{d}^\text{nc}(\bm{y}) + \grad_y V_\text{d}(\bm{y})$ is only non-zero in the entries corresponding to the driven particles and only depends on their coordinates, we can simplify this to
\begin{align}
\sigma_\text{st} &\leq \sigma_\text{st,d} \label{entropy-variational-driven} \\
&= \frac{1}{T} \inf_{V_\text{d}} \Big\langle \big( \bm{F}_\text{d}^\text{nc} + \grad_y V_\text{d} \big) \bm{\mu}_\text{d} \big( \bm{F}_\text{d}^\text{nc} + \grad_y V_\text{d} \big) \Big\rangle_{\text{d}} \n .
\end{align}
Here $\av{\ldots}_\text{d}$ denotes an average with respect to the marginal probability density of the driven particles, $p_\text{st,d}(\bm{y}) = \int d\bm{z} \ p_\text{st}(\bm{y},\bm{z})$.
Crucially, \eqref{entropy-variational-driven} only depends on quantities involving the driven degrees of freedom: The non-conservative force $\bm{F}_\text{d}^\text{nc}(\bm{y})$, the mobility matrix $\bm{\mu}_\text{d}$ and the statistics of the driven particles encoded in $p_\text{st,d}(\bm{y})$.
This implies that we can obtain an upper bound on the entropy production rate by only observing the driven degrees of freedom.
Moreover, the right-hand side of \eqref{entropy-variational-driven} corresponds precisely to the entropy production rate of the $M$ driven particles with steady state $p_\text{st,d}(\bm{y})$ in the absence of interactions with the passive particles.
This implies that any interaction between the driven and the passive particles necessarily decreases the entropy production rate compared to the case where there is no interaction.
At first glance, this appears to be counter-intuitive, since the interaction between the driven and the passive particles generally also induces currents among the passive particles, thereby increasing the dissipation.
However, the same interaction also reduces the response of the active particles to the driving force and therefore the corresponding current.
\eqref{entropy-variational-driven} implies that the second effect always outweighs the first, that is, the reduction in the currents of the driven particles is always larger than the currents induced in the passive particles.

The bound \eqref{entropy-variational-driven} is also useful from a practical point of view:
In many cases, there are only one or a few driven (or probe) particles, whose properties, including the non-conservative force driving them, are well-known.
By contrast, there may be many different species of passive particles, with complicated interactions among each other and with the driven particles.
The upper bound \eqref{entropy-variational-driven} is independent of these details, instead relying only on the properties and a measurement of the driven particles.
We stress that $p_\text{st,d}(\bm{y})$ is the probability density of the driven particles in the actual physical system (including interactions between driven and passive particles), and therefore, the average in \eqref{entropy-variational-driven} can be evaluated from observed trajectory data of the driven particles.
In complete analogy to \eqref{entropy-variational-nonconservative} we also have
\begin{align}
\sigma_\text{st} \leq \frac{1}{T} \Big\langle \big( \bm{F}_\text{d}^\text{nc} - \Av{\bm{F}_\text{d}^\text{nc}}_\text{d} \big) \bm{\mu}_\text{d} \big( \bm{F}_\text{d}^\text{nc} - \Av{\bm{F}_\text{d}^\text{nc}}_\text{d} \big) \Big\rangle_\text{d},
\end{align}
where, likewise, all quantities on the right-hand side only depend on the driven particles.
We also remark that, while in the above we referred to the driven and passive degrees of freedom as particles, \eqref{entropy-variational-driven} is not restricted to this case.
For example, the driven degrees of freedom may correspond to the spatial directions in which the non-conservative force acts, while the passive degrees of freedom are the spatial directions unaffected by the non-conservative force.

Even if many (or all) particles are driven, it is still possible to use \eqref{entropy-variational-nonconservative} obtain a useful upper bound on the entropy production by exploiting symmetries in the system.
For example, consider the case where the non-conservative force acting on each particles is the same and only depends on the coordinate of the respective particle, that is, $\bm{F}^\text{nc}(\bm{x}) = \sum_k \bm{F}^\text{nc}_k(\bm{x}_i)$.
Then we can choose a sum of one-body potentials $V(\bm{x}) = \sum_k V_k(\bm{x}_k)$, which yields the bound
\begin{align}
\sigma_\text{st} \leq \frac{1}{T} \sum_{k} \inf_{V_k} \Big\langle \big(\bm{F}_k^\text{nc} + \grad V_k \big) \bm{\mu}_k \big(\bm{F}_k^\text{nc} + \grad V_k \big)\Big\rangle_k,
\end{align}
where $\bm{\mu}_k$ is the mobility matrix of particle $k$.
The right-hand side is now written as a sum of one-particle expressions, which can be evaluated by using the one-particle probabilities $p_{\text{st},k}(\bm{x}_k)$.
In particular, if all particles are identical, then the above simplifies to
\begin{align}
\sigma_\text{st} \leq \frac{K}{T} \inf_{V} \Big\langle \big(\bm{F}_1^\text{nc} + \grad V \big) \bm{\mu}_1 \big(\bm{F}_1^\text{nc} + \grad V \big)\Big\rangle_1 \label{entropy-bound-one-particle} .
\end{align}
This implies that, for systems of identical, driven particles, the total entropy production is less than sum of one-particle entropy productions.
Similar to the case of driven and passive particles, dissipation is reduced by the interactions between the particles and the resulting correlations.
It is worth emphasizing again that $p_1(\bm{x}_1)$ is the one-particle density in the interacting system, and therefore the average in \eqref{entropy-bound-one-particle} can be directly evaluated from the trajectory data of the latter.

We remark that the finding that interactions reduce dissipation relies on a particular identification of the non-interacting system.
Let us write $\bm{F}(\bm{x}) = \sum_i \bm{F}_i(\bm{x}_i) - \grad U^\text{int}(\bm{x})$, with the interaction potential $U^\text{int}(\bm{x})$ and one-body forces $\bm{F}_i(\bm{x}_i) = - \grad_i U_i(\bm{x}_i) + \bm{F}_i^\text{nc}(\bm{x}_i)$.
We refer to this interacting out of equilibrium system as $S$.
In most cases, we would identify the non-interacting system $S_1$ as the one with $U^\text{int}(\bm{x}) = 0$, that is, a system of independent particles moving under the influence of only the one-body force.
This has the advantage that the solution can usually be obtained much easier than for the interacting system.
However, there is no definite general relation between the corresponding one-body entropy production $\sigma_{\text{st},1}$ of $S_1$ and the entropy production $\sigma_\text{st}$ of the interacting system $S$, so the former cannot be used to draw any conclusions about the latter.
This approach further has the downside that the statistics of the non-interacting system $S_1$ cannot be obtained from a measurement of $S$.
Since there is generally no way of just turning off the interactions, the non-interacting system $S_1$ may not be accessible in practice.
By contrast, in the bound \eqref{entropy-bound-one-particle}, we identify the non-interacting system $\bar{S}_1$ as the one with the same one-particle probability density $p_{\text{st},1}(\bm{x}_1)$ as the interacting system.
This has the obvious downside that we do not know the corresponding one-particle potential explicitly; the latter is defined by requiring that it has to reproduce the probability density $p_{\text{st},1}(\bm{x}_1)$ in the presence of the one-body non-conservative force $\bm{F}_1^\text{nc}(\bm{x})$.
The advantage of interpreting $\bar{S}_1$ as the non-interacting system is, however, that its one-body entropy production $\bar{\sigma}_{\text{st},1}$ always yields an upper bound on the entropy production on the entropy production of $S$, $\sigma_\text{st} \leq K \bar{\sigma}_{\text{st},1}$.
Further, $\bar{\sigma}_{\text{st},1}$ can be computed from the one-body probability density in the interacting system $S$---and thus from measured trajectory data of the latter---using the variational expression in \eqref{entropy-bound-one-particle}.
In conclusion, if we treat $S_1$ as the non-interacting system, then adding interactions between the particles may increase the dissipation, since doing so also changes the one-body density.
However, when comparing the interacting system $S$ with the non-interacting system $\bar{S}_1$ with the same one-body density, the interacting system always has reduced dissipation.
We note that $S_1$ and $\bar{S}_1$ coincide if the interactions do not change the one-body density.
This also implies that a necessary condition for interactions to increase dissipation is that they significantly change the one-body density compared to the non-interacting system $S_1$.
It would be interesting to see if this statement can be made more quantitative, that is, whether a direct relation between a potential increase in dissipation and the change in the one-body density can be obtained.

\section{Multiple heat baths} \label{sec-multiple-baths}
In the previous Section, we focused on the case where the entire system is in contact with a single equilibrium environment; in that case, any non-equilibrium effects stem from non-conservative forces.
By contrast, if a system is in contact with multiple heat baths at different temperatures, this generally induces heat flows and therefore drives the system out of equilibrium even if all forces are conservative.
For concreteness, let us consider $M$ particles in $n$ dimensions, $\bm{x} = (\bm{x}_1,\ldots,\bm{x}_M)$ with $\bm{x}_1 = (x_{1,1},\ldots,x_{1,n})$.
We allow each particle $\bm{x}_i$ to be in contact with a heat bath at a different temperature $T_i$.
We assume that the mobility matrix is block-diagonal, $\bm{\mu} = (\bm{\mu}_1,\ldots,\bm{\mu}_M)$, where $\bm{\mu}_i$ is the $n \times n$ mobility matrix of particle $\bm{x}_i$.
Since we are mainly interested in the effect of different heat baths on the dissipation, we restrict the discussion to the case of conservative forces.
We write the potential as $U(\bm{x}) = U^\text{int}(\bm{x}) + \sum_i U_i(\bm{x}_i)$, that is, we distinguish between the interaction potential $U^\text{int}(\bm{x})$ which may depend on all coordinates, and the one-body potentials $U_i(\bm{x}_i)$, which can be different for each particle.
From this and the preceding discussion, it is clear that the system is in equilibrium if all the temperatures are the same (in that case, we just have a conservative system in contact with an equilibrium environment) or if the interaction potential vanishes (in that case, we have $M$ independent equilibrium systems in contact with different equilibrium environments).
So, any possible entropy production stems from the interplay between interactions and the temperature differences between the heat baths.
The diffusion matrix for the current setup is the block-diagonal matrix $\bm{B}$ with entries $\bm{\mu}_i T_i$, and the local mean velocity has components
\begin{align}
\bm{\nu}_{\text{st},i}(\bm{x}) = - \bm{\mu}_i \big( \grad_i U^\text{int}(\bm{x}) + \grad_i U_i(\bm{x}_i) + T_i \grad_i \ln p_\text{st}(\bm{x}) \big),
\end{align}
where $\grad_i$ denotes the gradient with respect to $\bm{x}_i$.
In this case, \eqref{entropy-variational} is written as
\begin{align}
\sigma_\text{st} &= \sup_{\psi} \sum_i \\
&\qquad \Big\langle \big( \grad_i U^\text{int} + \grad_i U_i + T_i \grad_i \ln p_\text{st} - T_i \grad_i \psi \big) \nn
& \qquad \times \frac{\bm{\mu}_i}{T_i}  \big( \grad_i U^\text{int} + \grad_i U_i + T_i \grad_i \ln p_\text{st} - T_i \grad_i \psi \big) \Big\rangle \n .
\end{align}
We can absorb the gradient of the probability density and the one-body potentials into $\psi(\bm{x})$,
\begin{align}
\sigma_\text{st} &= \sup_{\psi} \sum_i \Big\langle \big( \grad_i U^\text{int} - T_i \grad_i \psi \big) \label{entropy-variational-heat-baths} \\
&\hspace{2cm} \times \frac{\bm{\mu}_i}{T_i}  \big( \grad_i U^\text{int} - T_i \grad_i \psi \big) \Big\rangle \n .
\end{align}
Rescaling $\psi(\bm{x})$ by a global constant $\alpha$ and minimizing with respect to $\alpha$ yields the equivalent expression
\begin{align}
\sigma_\text{st} &= \inf_{\psi} \bigg(   \Av{\grad U^\text{int} \bm{\mu} \bm{T}^{-1} \grad U^\text{int}} - \frac{ \Av{\grad U^\text{int} \bm{\mu} \grad \psi}^2}{\Av{\grad \psi \bm{\mu}\bm{T} \grad \psi}} \bigg) .
\end{align}
Where we defined the diagonal matrix $\bm{T}$ that has entries $T_i$ on the diagonal elements corresponding to $\bm{x}_i$.
We now choose $\psi(\bm{x}) = U^\text{int}(\bm{x})$, which results in the upper bound
\begin{align}
\sigma_\text{st} &\leq \bigg(   \Av{\grad U^\text{int} \bm{\mu} \bm{T}^{-1} \grad U^\text{int}} - \frac{ \Av{\grad U^\text{int} \bm{\mu} \grad U^\text{int}}^2}{\Av{\grad U^\text{int} \bm{\mu}\bm{T} \grad U^\text{int}}} \bigg) .
\end{align}
Writing this explicitly in terms of $\bm{\mu}_i$ and $\bm{T}_i$, we obtain
\begin{align}
\sigma_\text{st} &\leq \frac{\sum_{i,j} \frac{(T_i-T_j)^2}{T_i T_j} \mathcal{U}_i \mathcal{U}_j }{2 \sum_i T_i \mathcal{U}_i} \label{entropy-bound-heat-baths} \\
\text{with}\quad \mathcal{U}_i &= \Av{\grad_i U^\text{int} \bm{\mu}_i \grad_i U^\text{int}} \n .
\end{align}
The constant $\mathcal{U}_i$ quantifies the magnitude of the interaction force acting on particle $i$.
Note that each term in the numerator is positive and vanishes if either the temperature difference between two particles vanishes (pair of particles is at thermal equilibrium with respect to each other) or the interaction force on one of the particles vanishes identically (particle is independent from the remaining particles).
The bound \eqref{entropy-bound-heat-baths} allows us to obtain an upper estimate on the entropy production rate in heat conduction systems, which depends only on the temperature differences and the interaction forces---it only contains contributions from pairs of particles between which a heat flow can occur in principle.
However, we remark that, for example for a linear chain of $M$ particles, the bound \eqref{entropy-bound-heat-baths} becomes trivial in the continuum limit of vanishing temperature differences between neighboring particles and increasing particle number: 
The right-hand side is proportional to $M$, whereas the left-hand side is expected to be of order $1$.
The reason is that we allowed for arbitrary interactions, which means that particles at opposite ends of the chain, where the temperature difference is not small, could in principle interact.
It may be possible to obtain a more informed bound that explicitly takes into account the spatial structure of the interactions, such as nearest neighbor or finite-range interactions.
We leave this problem for future research, noting that \eqref{entropy-bound-heat-baths} is expected to be most useful for systems with few degrees of freedom.

\section{Spatial temperature variations} \label{sec-temp-gradient}
Finally, we also allow the temperature and/or mobility to depend on the coordinates $\bm{x}$ themselves.
For simplicity, we restrict the discussion to the case where all degrees of freedom are in contact with the same, spatially inhomogeneous heat bath at temperature $T(\bm{x})$.
As has been shown in Ref.~\cite{Jay95}, the appropriate form of the local mean velocity is in this case
\begin{align}
\bm{\nu}_\text{st}(\bm{x}) = \bm{\mu}(\bm{x}) \big( - \grad U(\bm{x}) - \grad T(\bm{x}) - T(\bm{x}) \grad \ln p_\text{st}(\bm{x}) \big) .
\end{align}
From \eqref{entropy-variational} we then have
\begin{align}
\sigma_\text{st} &= \inf_\psi \Big\langle \big( \grad U + \grad T + T \grad \ln p_\text{st} - T \grad \psi \big) \\
&\qquad \times \frac{\bm{\mu}}{T} \big( \grad U + \grad T + T \grad \ln p_\text{st} - T \grad \psi \big) \Big\rangle \n .
\end{align}
As before, we can absorb the term involving the gradient of the probability density into $\psi(\bm{x})$,
\begin{align}
\sigma_\text{st} &= \inf_\psi \Big\langle \big( \grad U + \grad T - T \grad \psi \big) \\
&\qquad \times \frac{\bm{\mu}}{T} \big( \grad U + \grad T - T \grad \psi \big) \Big\rangle \n .
\end{align}
We write $\psi(\bm{x}) = (U(\bm{x}) - U_0)/T(\bm{x}) + \ln T(x) + \phi(\bm{x})$ with an arbitrary constant $U_0$,
\begin{align}
\sigma_\text{st} &= \inf_\phi \bigg\langle \bigg( \frac{U - U_0}{T} \grad T - T \grad \phi \bigg) \\
&\qquad \times \frac{\bm{\mu}}{T} \bigg( \frac{U - U_0}{T} \grad T  - T \grad \phi \bigg) \bigg\rangle \n .
\end{align}
Again rescaling $\phi(\bm{x})$ by a constant factor $\alpha$ and minimizing with respect to $\alpha$, we obtain
\begin{align}
\sigma_\text{st} &= \inf_\phi \Bigg( \Av{ \bigg( \frac{U - U_0}{T} \bigg)^2 \grad T \frac{\bm{\mu}}{T} \grad T } \\
& \hspace{2cm} -\frac{\Av{\big( \frac{U - U_0}{T} \big) \grad T \bm{\mu} \grad \phi}^2}{\Av{\grad \phi \bm{\mu} T \grad \phi}} \Bigg) \n .
\end{align}
Finally, choosing $\phi(\bm{x}) = \ln T(\bm{x})$, we obtain the upper bound
\begin{align}
\sigma_\text{st} &\leq \Av{ \bigg( \frac{U - U_0}{T} \bigg)^2 \grad T \frac{\bm{\mu}}{T} \grad T } -\frac{\Av{\big( \frac{U - U_0}{T} \big) \grad T \frac{\bm{\mu}}{T} \grad T}^2}{\Av{\grad T \frac{\bm{\mu}}{T} \grad T}} .
\end{align}
The right-hand side vanishes if either the temperature is homogeneous, $\grad T(\bm{x}) = 0$, or if the potential and temperature profile are linearly related, $U(\bm{x}) = c T(\bm{x}) + U_0$.
In general, the upper bound depends on the relative magnitude of the temperature gradient $\mathcal{T}(\bm{x}) = \grad T(\bm{x}) \bm{\mu} \grad T(\bm{x})/T(\bm{x})$ and the ratio $\mathcal{R}(\bm{x}) = (U(\bm{x}) - U_0)/T(\bm{x})$ between the potential energy and the temperature.
In terms of these quantities we have,
\begin{align}
\sigma_\text{st} &\leq \Av{\mathcal{T}} \Bigg( \frac{\Av{\mathcal{R}^2 \mathcal{T}}}{\Av{\mathcal{T}}} - \bigg( \frac{\Av{\mathcal{R} \mathcal{T}}}{\Av{\mathcal{T}}} \bigg)^2 \Bigg) \label{entropy-bound-temp-gradient} .
\end{align}
We note that the second factor is equal to the variance of $\mathcal{R}(\bm{x})$ with respect to the probability density $q(\bm{x}) = \mathcal{T}(\bm{x}) p_\text{st}(\bm{x})/\Av{\mathcal{T}}$.
If the ratio $\mathcal{R}(\bm{x})$ between potential energy and temperature is bounded, $\mathcal{R}^\text{min} \leq \mathcal{R}(\bm{x}) \leq \mathcal{R}^\text{max}$ with $\Delta \mathcal{R} = \mathcal{R}^\text{max} - \mathcal{R}^\text{min}$, then we can further bound this variance from above,
\begin{align}
\sigma_\text{st} &\leq \frac{\Delta \mathcal{R}^2}{4}\Av{\mathcal{T}} .
\end{align}
If the magnitude of the temperature gradient is also bounded, $\mathcal{T}(\bm{x}) \leq \mathcal{T}^\text{max}$, then we further have
\begin{align}
\sigma_\text{st} &\leq \frac{\Delta \mathcal{R}^2}{4} \mathcal{T}^\text{max} \label{entropy-bound-temp-gradient-simple} .
\end{align}
Note that the above still depends on the global shift $U_0$ of the potential, which is in principle arbitrary, but should be chosen such as to minimize the range $\Delta \mathcal{R}$ in order to obtain a tight bound.
One heuristic way of doing so is to plot $U(\bm{x})$ as a function of $T(\bm{x})$ and then perform a linear regression on the data.
If a linear relationship of the form $U(\bm{x}) = c T(\bm{x}) + U_0$ exists, then the parameter $U_0$ obtained from the fit will lead to a constant ratio $\mathcal{R}(\bm{x})$ and thus $\Delta \mathcal{R} = 0$.
In other cases, we obtain a linear approximation of the relation between $U(\bm{x})$ and $T(\bm{x})$ and the parameter $U_0$ obtained from the fit will generally result in a tighter bound than simply choosing $U_0 = 0$.
We stress that this procedure only requires knowledge of the spatial dependence of $T(\bm{x})$ and $U(\bm{x})$ and the bound can be computed without solving the dynamics.
Thus, \eqref{entropy-bound-temp-gradient-simple} provides a simple and explicit upper bound on the entropy production rate in terms of the parameters of the model.

\section{Periodic boundary conditions} \label{sec-periodic}
When introducing the orthogonality relation \eqref{orthogonality}, we assumed natural or reflecting boundary conditions, which, in particular, implies that the probability current vanishes at the boundaries and we can thus neglect the boundary terms when integrating by parts.
For systems with periodic boundary conditions, on the other hand, there is generally a finite probability current across the boundary.
Let us consider the expression
\begin{align}
\Av{\grad \psi \bm{\nu}_\text{st}} &= \int_{\Omega} d\bm{x} \ p_\text{st}(\bm{x}) \bm{\nu}_\text{st}(\bm{x}) \grad \psi(\bm{x}) \\
&= p_\text{st}(\bm{x}) \bm{\nu}_\text{st}(\bm{x}) \grad \psi(\bm{x}) \Big\vert_{\partial \Omega} \nn
& \hspace{1cm} - \int_{\Omega} d\bm{x} \ \psi(\bm{x}) \grad \big( p_\text{st}(\bm{x}) \bm{\nu}_\text{st}(\bm{x}) \big) \n ,
\end{align}
where $\Omega$ denotes the configuration space and $\partial \Omega$ its boundary.
The second term vanishes due to \eqref{fpe-steady}.
However, the first term only vanishes if the function $\psi(\bm{x})$ has the same periodicity as the probability current $\bm{\nu}_\text{st}(\bm{x}) p_\text{st}(\bm{x})$, which is a periodic function in the directions where the boundary conditions are periodic.
Thus, the orthogonality relation \eqref{orthogonality} only holds for gradients of periodic functions $\psi(\bm{x})$, and the same is true for all gradient fields used in subsequent expressions.
As an example, suppose that the system is periodic with periodicity $L_1$ in direction $x_1$ and satisfies natural boundary conditions in directions $x_2$ and $x_3$.
This is the case for a particle moving in linear channel in three dimensions, where $x_1$ is the direction along the channel and the particle is assumed to be confined in the transverse directions.
In this case, we require the function $\psi(x_1,x_2,x_3)$ to be periodic in direction $x_1$, $\psi(x_1+L,x_2,x_3) = \psi(x_1,x_2,x_3)$, whereas in directions $x_2$ and $x_3$ the only requirement is that the function does not grow too fast such that $\psi(\bm{x}) \bm{\nu}_\text{st}(\bm{x}) p_\text{st}(\bm{x})$ vanishes as $|x_2|,|x_3| \rightarrow \infty$ and the integral over these directions is finite.
For the variational expression \eqref{entropy-variational} and results derived from it, we thus have to replace the minimization over all gradient fields with the minimization over gradients of functions with the appropriate periodicity.
This also implies that any specific choice of $\psi(\bm{x})$ that is used to obtain an upper bound should satisfy the same periodicity constraints.
Since the potential and temperature profile have to be periodic functions to ensure the periodicity of the steady state density, \eqref{entropy-bound-heat-baths} and \eqref{entropy-bound-temp-gradient} also hold for systems with periodic boundary conditions.
However, the choice $\psi(\bm{x}) = \bm{x} \Av{\bm{F}^\text{nc}}$ used to obtain \eqref{entropy-bound-nonconservative} is not a periodic function and therefore not admissible, and \eqref{entropy-variational-nonconservative} does not hold for periodic systems.
A simple counterexample is a particle in a one-dimensional periodic potential $U(x+L) = U(x)$, driven by a constant bias $F^\text{nc}(x) = F_0$.
Here, the variance of the bias force is zero, yet the system is out of equilibrium and has a finite entropy production rate.
Similarly, \eqref{entropy-bound-constant} does not hold, because in this case, the effect of the bias force is largest if it acts over the entire length of the potential.
We may use $\psi(\bm{x}) = \bm{x}_\text{np} \Av{\bm{F}^\text{nc}}$, where the vector $\bm{x}_\text{np}$ is equal to $\bm{x}$ in the non-periodic directions and equal to $0$ (and thus trivially periodic) in the periodic directions.
Then, we obtain the bound
\begin{align}
\sigma_\text{st} &\leq \frac{1}{T} \Big( \Av{\bm{F}^\text{nc}_\text{p} \bm{\mu} \bm{F}^\text{nc}_\text{p}} \\
&\hspace{1cm}  + \Av{\big(\bm{F}^\text{nc}_{\text{np}} - \Av{\bm{F}^\text{nc}_\text{np}} \big) \bm{\mu} \big(\bm{F}^\text{nc}_{\text{np}} - \Av{\bm{F}^\text{nc}_\text{np}} \big)} \Big) \n .
\end{align}
Here, $\bm{F}^\text{nc}_\text{p}(\bm{x})$ is the non-conservative force acting in the periodic directions, while $\bm{F}^\text{nc}_\text{np}(\bm{x})$ is the non-conservative force acting in the non-periodic directions.

\section{Underdamped dynamics} \label{sec-underdamped}
So far, we always assumed that the system is described by an overdamped Langevin dynamics of the type \eqref{langevin}.
This assumption is well-justified for small particles diffusing in a relatively dense viscous environment, whose thermal relaxation time, $\tau_\text{th} = \frac{\gamma}{m}$ with $\gamma$ the friction coefficient and $m$ the particle mass, is typically orders of magnitude shorter than the characteristic timescales of the motion in the force field.
However, for large particles or dilute environments, inertia and thermal relaxation can no longer be neglected and we need to describe the dynamics using the underdamped Langevin equations
\begin{align}
\dot{\bm{x}}(t) &= \bm{v}(t) \label{langevin-underdamped}\\
\bm{m} \dot{\bm{v}}(t) &= \bm{F}(\bm{x}(t)) - \bm{\gamma} \bm{v}(t) + \sqrt{2 \bm{\gamma} T} \bm{\xi}(t) \n .
\end{align}
Here $\bm{v}$ are the velocities of the particles, $\bm{m}$ is a diagonal matrix containing the masses of the individual particles and $\bm{\gamma}$ is the matrix of friction coefficients, which we assume to be block-diagonal $\bm{\gamma} = (\bm{\gamma}_1,\ldots,\bm{\gamma}_K)$, where $\bm{\gamma}_k$ is the symmetric and positive definite friction coefficient matrix of the $k$-th particle.
This form allows for different (and possibly anisotropic) friction forces for the different particles, however, we assume that the friction force does not induce any coupling between different particles.
Note that since we assume the matrix $\bm{m}$ to have the same entry $m_k$ on the diagonal elements corresponding to particle $k$, the matrices $\bm{\gamma}$ and $\bm{m}$ commute.
If the force is conservative, $\bm{F}(\bm{x}) = - \grad U(\bm{x})$, then the steady state of the system is in equilibrium described by the Boltzmann-Gibbs density,
\begin{align}
p_\text{st}(\bm{x},\bm{v}) &= \frac{\exp \big( - \frac{U(\bm{x})}{T} \big)}{\int d\bm{x} \ \exp \big( - \frac{U(\bm{x})}{T} \big)} \\
&\hspace{0.5cm} \times \frac{1}{(2 \pi T)^{\frac{N}{2}} \sqrt{\det(\bm{m})}} \exp \bigg( -\frac{\bm{v} \bm{m}^{-1} \bm{v} }{2 T} \bigg) . \n
\end{align}
Here $N$ is the total number of degrees of freedoms ($N = 3K$ for $K$ particles in three dimensions) and $\det$ denotes the determinant.
By contrast, for non-conservative forces $\bm{F}(\bm{x}) = - \grad U(\bm{x}) + \bm{F}^\text{nc}(\bm{x})$, the system is in a non-equilibrium steady state and the probability density no longer factorizes into a position and velocity-dependent part.
Instead, we have to solve the steady-state Klein-Kramers-Fokker-Planck equation,
\begin{align}
0 = \bigg( - \bm{v} \grad_x - &\frac{1}{\bm{m}} \grad_v \Big( \bm{F}(\bm{x}) \label{kfp} \\
& - \bm{\gamma} \bm{v} - \frac{\bm{\gamma} T}{\bm{m}} \grad_v \Big) \bigg) p_\text{st}(\bm{x},\bm{v}) \n .
\end{align}
The entropy production rate in the non-equilibrium steady state is given by the magnitude of the irreversible probability currents as
\begin{align}
\sigma_\text{st} = \frac{1}{T} \Av{\bigg(\bm{v} + \frac{T}{\bm{m}} \grad_v \ln p_\text{st} \bigg) \bm{\gamma} \bigg(\bm{v} + \frac{T}{\bm{m}} \grad_v \ln p_\text{st} \bigg)}_\text{st} \label{entropy-under} .
\end{align}
Here, we take a division by the matrix $\bm{m}$ to mean a multiplication by $\bm{m}^{-1}$.
In order to proceed, we define the steady-state position density and local mean velocity
\begin{align}
p_\text{st}^x(\bm{x}) &= \int d\bm{v} \ p_\text{st}(\bm{x},\bm{v}) \\
\bm{\nu}_\text{st}(\bm{x}) &= \frac{\int d\bm{v} \ \bm{v} p_\text{st}(\bm{x}, \bm{v})}{p_\text{st}^x(\bm{x})} . \n
\end{align}
Integrating \eqref{kfp} over $\bm{v}$, we see that these two quantities satisfy a continuity equation
\begin{align}
0 = - \grad_x \big( \bm{\nu}_\text{st}(\bm{x}) p_\text{st}^x(\bm{x}) \big) \label{continuity-underdamped} ,
\end{align}
in complete analogy to the overdamped case \eqref{fpe-steady}.
Since the local mean velocity is the average velocity conditioned on the position $\bm{x}$, we introduce the fluctuations of the velocity $\bm{v}$ around its local average,
\begin{align}
\bm{\omega}(\bm{x},\bm{v}) = \bm{v} - \bm{\nu}_\text{st}(\bm{x}) .
\end{align}
Using this, we can write \eqref{entropy-under} as
\begin{align}
\sigma_\text{st} &= \sigma_\text{st}^\text{lm} + \sigma_\text{st}^\text{vf} \qquad \text{with} \label{entropy-decomposition-under} \\
\sigma_\text{st}^\text{lm} &= \frac{1}{T} \Av{\bm{\nu}_\text{st} \bm{\gamma} \bm{\nu}_\text{st}}_\text{st} \qquad \text{and} \nn
\sigma_\text{st}^\text{vf} &= \frac{1}{T}\Av{\bigg(\bm{\omega} + \frac{T}{\bm{m}} \grad_v \ln p_\text{st} \bigg) \bm{\gamma} \bigg(\bm{\omega} + \frac{T}{\bm{m}} \grad_v \ln p_\text{st} \bigg)}_\text{st} . \n
\end{align}
The first term is formally similar to the overdamped entropy production rate \eqref{entropy} in that it measures the dissipation caused by the local mean flows. It is positive whenever the local mean velocity is non-zero, and we define it as the local mean (lm) contribution $\sigma_\text{st}^\text{lm}$.
The second term, on the other hand, measures the non-thermal velocity fluctuations around the local mean. It vanishes only when the velocity fluctuations are given by
\begin{align}
p_\text{st}(\bm{v} \vert \bm{x}) = \frac{1}{(2 \pi T)^{\frac{N}{2}} \sqrt{\det(\bm{m})}} \exp \bigg( -\frac{\bm{\omega} \bm{m}^{-1} \bm{\omega} }{2 T} \bigg) ,
\end{align}
where $p_\text{st}(\bm{v} \vert \bm{x}) = p_\text{st}(\bm{x},\bm{v})/p_\text{st}^x(\bm{x})$ is the conditional velocity density.
We call this contribution the velocity fluctuation (vf) contribution $\sigma_\text{st}^\text{vf}$.
We note that the latter can equivalently be written as
\begin{align}
\sigma_\text{st}^\text{vf} = \frac{1}{T} \Big( \Av{\bm{\omega} \bm{\gamma} \bm{\omega}}_\text{st} - T \text{tr}\big(\bm{m}^{-1} \big) \Big),
\end{align}
where tr denotes the trace of a matrix.
Just as in the overdamped case, we can also express the entropy production in terms of the work done by the non-conservative force,
\begin{align}
\sigma_\text{st} = \frac{1}{T} \Av{\bm{F}^\text{nc} \bm{v}}_\text{st} = \frac{1}{T} \Av{\bm{F}^\text{nc} \bm{\nu}_\text{st}}_\text{st},
\end{align}
where we used that $\bm{F}^\text{nc}(\bm{x})$ only depends on the position, so we can replace the velocity by the local mean value under the average.
Due to \eqref{continuity-underdamped}, we can add an arbitrary gradient field in the above expression, since the average of the latter vanishes,
\begin{align}
\sigma_\text{st} = \frac{1}{T} \Av{\big(\bm{F}^\text{nc} + \grad_x V\big) \bm{\nu}_\text{st}}_\text{st} .
\end{align}
We now use the Cauchy-Schwarz inequality and identify the local mean entropy production rate,
\begin{align}
\sigma_\text{st}^2 &\leq \frac{1}{T^2} \Av{\bm{\nu}_\text{st} \bm{\gamma} \bm{\nu}_\text{st}} \Av{\big(\bm{F}^\text{nc} + \grad_x V\big) \bm{\gamma}^{-1} \big(\bm{F}^\text{nc} + \grad_x V\big)}_\text{st} \nn
&= \sigma_\text{st}^\text{lm} \frac{1}{T}  \Av{\big(\bm{F}^\text{nc} + \grad_x V\big) \bm{\gamma}^{-1} \big(\bm{F}^\text{nc} + \grad_x V\big)}_\text{st} \label{entropy-bound-lm} .
\end{align}
Since the entropy production rate is given by the sum of two positive terms, see \eqref{entropy-decomposition-under}, we have $\sigma_\text{st}^\text{lm} \leq \sigma_\text{st}$ and therefore,
\begin{align}
\sigma_\text{st} \leq \inf_V \frac{1}{T}  \Av{\big(\bm{F}^\text{nc} + \grad_x V\big) \bm{\gamma}^{-1} \big(\bm{F}^\text{nc} + \grad_x V\big)}_\text{st} \label{entropy-variational-under} .
\end{align}
The right-hand side is precisely the same as in the overdamped variational expression \eqref{entropy-variational-nonconservative}, with the mobility matrix $\bm{\mu} = \bm{\gamma}^{-1}$.
While in the overdamped case, the minimization with respect to the potential $V(\bm{x})$ results in the exact entropy production rate, the corresponding expression in the underdamped case is an upper bound on the entropy production rate.
Crucially, however, any upper bound obtained using a specific choice of $V(\bm{x})$ is an upper bound in both cases.
This implies that we can use \eqref{entropy-variational-nonconservative} to obtain upper bounds on the entropy production rate, irrespective of whether the dynamics are over- or underdamped.
In particular, the findings of Section~\ref{sec-thermal} with regards to bounds on the entropy production in terms of one-particle statistics remain true for the underdamped case, and we can therefore apply them even if we are not sure whether the effects of inertia can be neglected for the particular system of interest.
Let us define the right-hand side of \eqref{entropy-variational-under} as $\hat{\sigma}_\text{st}$, that is, the upper estimate on $\sigma_\text{st}$ obtained by solving the minimization problem.
Intriguingly, this can be used together with the actual value of the entropy production rate to estimate the velocity-fluctuation contribution from \eqref{entropy-bound-lm},
\begin{align}
\frac{\sigma_\text{st}^\text{vf}}{\sigma_\text{st}} \leq 1 - \frac{\sigma_\text{st}}{\hat{\sigma}_\text{st}} \label{velocity-entropy-bound} .
\end{align}
We can estimate the contribution of the non-thermal velocity fluctuations to the overall entropy production by how much the upper bound exceeds the actual value of the latter.
In the overdamped limit, the right-hand side vanishes and thus the velocity fluctuations become thermal.

The above generalizes in a straightforward way to the case of multiple heat baths discussed in Section~\ref{sec-multiple-baths}.
Introducing the diagonal matrix $\bm{T}$, which contains the temperatures of the heat baths to which the individual particles are coupled, we have
\begin{align}
\sigma_\text{st} &= \sigma_\text{st}^\text{lm} + \sigma_\text{st}^\text{vf} \qquad \text{with} \\
\sigma_\text{st}^\text{lm} &= \Av{\bm{\nu}_\text{st} \bm{\gamma} \bm{T}^{-1} \bm{\nu}_\text{st}}_\text{st} \qquad \text{and} \nn
\sigma_\text{st}^\text{vf} &= \Av{\bigg(\bm{\omega} + \frac{\bm{T}}{\bm{m}} \grad_v \ln p_\text{st} \bigg) \bm{\gamma} \bm{T}^{-1} \bigg(\bm{\omega} + \frac{\bm{T}}{\bm{m}} \grad_v \ln p_\text{st} \bigg)}_\text{st} , \n
\end{align}
so \eqref{entropy-decomposition-under} generalizes in a straightforward way to the cases of several heat baths.
Here we assumed that either, the temperature associated with each spatial direction of a single particle is the same, or, that the friction matrix $\bm{\gamma}$ is diagonal, such that $\bm{\gamma}$ and $\bm{T}$ commute.
Again, we can equivalently express the entropy production rate in terms of the work done by the total force,
\begin{align}
\sigma_\text{st} = -\Av{ \grad_x U \bm{T}^{-1} \bm{\nu}_\text{st}}_\text{st} \label{entropy-force-underdamped} ,
\end{align}
where $U(\bm{x})$ contains both the one-body and interaction potentials.
In this case, we can add a term of the form $\bm{T} \grad_x \psi(\bm{x})$ to the force without changing the result,
\begin{align}
\sigma_\text{st} = -\Av{ \big(\grad_x U - \bm{T} \grad_x \psi \big) \bm{T}^{-1} \bm{\nu}_\text{st}}_\text{st} ,
\end{align}
since we still have $\Av{\grad \psi \bm{\nu}_\text{st}}_\text{st} = 0$ from \eqref{continuity-underdamped}.
As discussed in Section~\ref{sec-multiple-baths}, the one-body potentials can be absorbed into the gradient term, and we therefore can write the upper bound as
\begin{align}
\sigma_\text{st} \leq \inf_\psi \big\langle \big(\grad_x U^\text{int} & - \bm{T} \grad_x \psi \big) (\bm{\gamma} \bm{T})^{-1} \\
& \big(\grad_x U^\text{int} - \bm{T} \grad_x \psi \big) \big\rangle_\text{st} \n .
\end{align}
This reproduces \eqref{entropy-variational-heat-baths}, except that the minimization now generally yields an upper bound instead of the exact entropy production rate.
Nevertheless, the upper bound \eqref{entropy-bound-heat-baths} remains valid for underdamped dynamics.

Finally, for a spatially inhomogeneous temperature $T(x)$ and friction coefficient $\bm{\gamma}(x)$ (for simplicity we assume $T(\bm{x})$ to be scalar-valued in the following), we have
\begin{align}
\sigma_\text{st} &= \sigma_\text{st}^\text{lm} + \sigma_\text{st}^\text{vf} \qquad \text{with} \\
\sigma_\text{st}^\text{lm} &= \Av{\frac{1}{T} \bm{\nu}_\text{st} \bm{\gamma} \bm{\nu}_\text{st}}_\text{st} \qquad \text{and} \nn
\sigma_\text{st}^\text{vf} &= \Av{\frac{1}{T} \bigg(\bm{\omega} + \frac{T}{\bm{m}} \grad_v \ln p_\text{st} \bigg) \bm{\gamma} \bigg(\bm{\omega} + \frac{T}{\bm{m}} \grad_v \ln p_\text{st} \bigg)}_\text{st} . \n
\end{align}
Compared to \eqref{entropy-decomposition-under}, the only difference is that the temperature now has to be taken inside the average.
We can again obtain an equivalent expression in terms of the force, which now, however, acquires an additional term proportional to the temperature gradient,
\begin{align}
\sigma_\text{st} = - \Av{\frac{1}{T} \grad_x U \bm{\nu}_\text{st}}_\text{st} - \Av{\frac{\bm{v} \bm{m} \bm{v}}{2 T} \bm{v} \grad_x \ln T}_\text{st} .
\end{align}
The presence of the second term prohibits us from directly extending the results of Section~\ref{sec-temp-gradient} to the underdamped case.
It has been previously found that underdamped systems in the presence of a temperature gradient exhibit a type of \enquote{hidden entropy production} \cite{Cel12}, which stems from the continuous thermalization of the velocity degrees of freedom as the particle moves through the temperature gradient.
This contribution to the dissipation does not vanish in the overdamped limit and cannot be estimated from the statistics of the position alone.
Therefore, a bound of the form \eqref{entropy-bound-temp-gradient}, which only relies on a measurement of the position, cannot be expected to reproduce this contribution.

\section{Demonstration} \label{sec-demonstration}

\subsection{Interacting particles on a ring} \label{sec-demonstration-interacting}
\begin{figure}
\includegraphics[width=0.47\textwidth, trim = 4cm 4cm 2cm 1cm, clip]{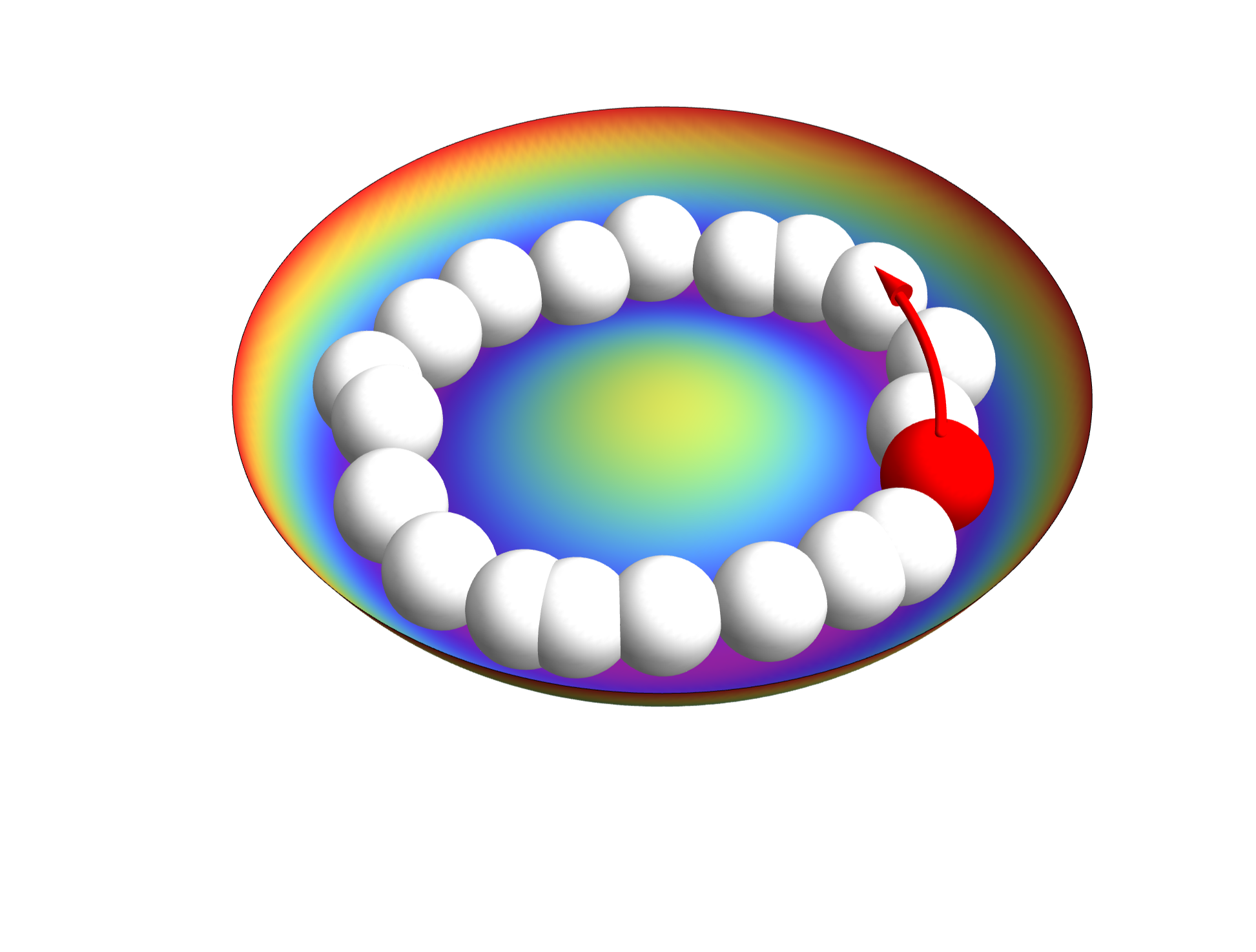}
\caption{Illustration of the interacting particle system discussed in Section \ref{sec-demonstration-interacting}. The orange surface represents the trapping potential, in which driven (red, in this case one) and passive (white, in this case 19) particles diffuse and interact. The red arrow indicates the non-conservative driving force. The radius of the particles is chosen as the length scale $\lambda$ of the interaction potential. } \label{fig-interacting-potential}
\end{figure}

In order to investigate the tightness and usefulness of the bounds derived in Section \ref{sec-thermal}, we study a system of $N$ Brownian particles with coordinates $\bm{x} = (\bm{x}_1,\ldots,\bm{x}_N)$, where $\bm{x}_i = (x_{i,1},x_{i,2})$ in two dimensions.
The particles are trapped inside the radially symmetric one-body potential
\begin{align}
U_1(r_1) = U_0 \bigg( \frac{1}{4} \frac{r_1^4}{L^4} - \frac{\alpha}{2} \frac{r_1^2}{L^2} \bigg),
\end{align}
where $r_1 = \Vert \bm{x}_1 \Vert$ is the distance of a particle from the origin.
For $\alpha > 0$, this potential has a maximum at $r_1 = 0$ and a minimum at $r_1 = L \sqrt{\alpha}$, resulting in a ring-shaped groove (see Fig.~\ref{fig-interacting-potential}).
The particles interact via the truncated and shifted Lennard-Jones-type two-body potential,
\begin{gather}
U_{12}(r_{12}) = \left\lbrace \begin{array}{ll}
U^\text{LJ}(r_{12}) - U^\text{LJ}(r_\text{c})) & \text{for} \; r_{12} < r_\text{c} \\ 0 &\text{for} \; r_{12} \geq r_\text{c}
\end{array} \right.  \\
 \text{with} \qquad U^\text{LJ}(r) = U_\text{I} \Bigg( \bigg( \frac{\lambda}{r} \bigg)^{2 \delta} - 2 \bigg( \frac{\lambda}{r} \bigg)^{\delta} \bigg) \Bigg) . \n 
\end{gather}
Here, $\lambda$ is the characteristic length scale of the interactions, the exponent $\delta > 0$ determines the steepness of the potential and $r_\text{c}$ is a cutoff length.
$r_{12} = \Vert \bm{x}_1 - \bm{x}_2 \Vert$ is the distance between two particles.
Note that the potential is repulsive for $r < \lambda$ and attractive for $\lambda < r < r_\text{c}$; for $r_\text{c} = \lambda$, the potential is purely repulsive.
The total potential is given by $U(\bm{x}) = \sum_{i} U_1(r_i) + \sum_{i \neq j} U_{12}(r_{ij})$.
The environment is at temperature $T$ and we take the mobility to be the same constant $\mu$ for each particle.
In addition, we introduce the non-conservative one-body force,
\begin{align}
\bm{F}_1^\text{nc}(\bm{x}_1) = \frac{\mathcal{F}}{\Vert \bm{x}_1 \Vert} \begin{pmatrix} x_{1,2} \\ -x_{1,1} \end{pmatrix},
\end{align} 
which corresponds to a constant-magnitude force $\Vert \bm{F}_1^\text{nc} \Vert = \mathcal{F}$ driving a particle in a circle around the origin.
Depending on the situation, this non-conservative force acts on only one particle $\bm{F}^\text{nc}(\bm{x}) = \bm{F}_1^\text{nc}(\bm{x}_1)$), some particles, or all particles ($\bm{F}^\text{nc}(\bm{x}) = \sum_i \bm{F}_1^\text{nc}(\bm{x}_i)$).
In either case, since the system is radially symmetric, we have $\Av{\bm{F}^\text{nc}} = 0$, and so the bound \eqref{entropy-bound-nonconservative} simplifies to
\begin{align}
\sigma_\text{st} \leq \frac{\mu K \mathcal{F}^2}{T} = \sigma_\text{st,1d}, \label{dem1-upper-bound}
\end{align}
where $K$ is the number of driven particles.
This bound only depends on a small number of parameters of the system, namely, the mobility, the temperature, the strength of the driving and the number of driven particles.
In particular, the bound is independent of the precise nature of the interaction or trapping potentials, but only depends on the 
Note that the same bound is obtained if the driven and passive particles have different physical properties, e.~g.~mobility.
In this case, the bound is already the optimal one that can be obtained using the one-particle statistics of the driven particles, see \eqref{entropy-bound-one-particle},
\begin{align}
\sigma_\text{st,1d} = \frac{K}{T} \inf_{V} \Big\langle \big(\bm{F}_1^\text{nc} + \grad V \big) \bm{\mu}_1 \big(\bm{F}_1^\text{nc} + \grad V \big)\Big\rangle_1 .
\end{align}
To see this, we note that the entire problem is radially symmetric, and so the same has to be true for the optimal one-body potential $V(\bm{x}_k)$, where $k$ denotes one of the driven particles.
Then, the gradient of $V(\bm{x}_k)$ is along the radial direction and, therefore, perpendicular to the non-conservative force $\bm{F}_1^\text{nc}(\bm{x}_k)$.
Consequently, the above expression is minimized for $\grad_k V(\bm{x}_k) = 0$, which yields \eqref{dem1-upper-bound}, noting that the magnitude of the driving force is constant.

\begin{figure}
\includegraphics[width=0.47\textwidth]{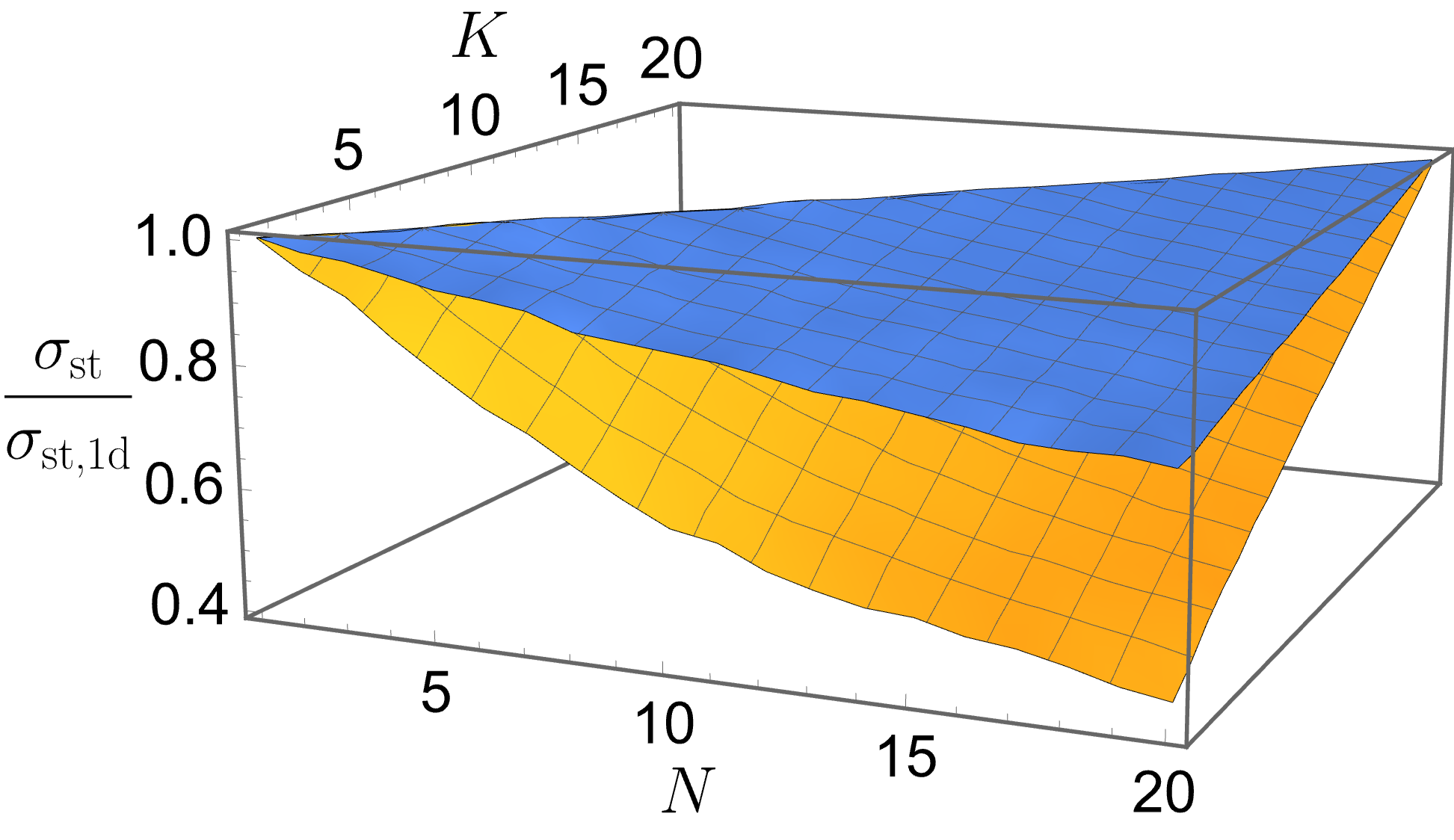}
\caption{Ratio of the actual entropy production rate $\sigma_\text{st}$ to the upper bound $\sigma_\text{st,1d}$ (\eqref{dem1-upper-bound}) obtained from the one-particle statistics of the driven particles, as a function of the total particle number $N$ and the number of driven particles $K$. The orange surface is the data for both repulsive and attractive interaction ($r_c = 2 \lambda$), the blue surface for only repulsive interaction ($r_c = \lambda$). } \label{fig-interacting-entropy}
\end{figure}
We now compare the upper bound \eqref{dem1-upper-bound} to the actual entropy production.
The latter we compute from numerical simulations of the system outlined above.
Concretely, since, as discussed in Section \ref{sec-underdamped}, the bounds also apply to underdamped dynamics, we consider particles with mass $m = 0.5$ in contact with a heat bath at temperature $T = 1.0$ and with friction coefficient $\gamma = 1$ (i.~e.~$\mu = 1/\gamma = 1$).
These particles evolve according to \eqref{langevin-underdamped} with the force given by
\begin{align}
\bm{F}(\bm{x}) = - \sum_{i = 1}^N \grad_i U_1(r_i) &- \frac{1}{2} \sum_{i,j = 1 \vert i \neq j}^N \grad_i U_{12}(r_{ij}) \\
& - \sum_{i = 1}^K \bm{F}_1^\text{nc}(\bm{x}_i) \n .
\end{align}
We set the parameters of the trapping potential to $U_0 = 10$, $L = 10$ and $\alpha = 1$, those of the interaction potential to $U_\text{I} = 4$, $\lambda = 2$ and $\delta = 2$, and the strength of the driving force to $\mathcal{F} = 1$.
Note that for these parameters, the depth of both the groove of the trapping potential ($2.5 T$) and the minimum of the interaction potential ($1.125 T$) is comparable to the temperature.
We further introduce a short-range cutoff $r_m = \lambda/3$ and replace the interaction force by its maximal value at $r_{ij} = r_m$ for shorter distances.
The reason for doing so is that it allows us to choose a more reasonable size for the time step in the simulations, since it prevents the interaction force from becoming arbitrarily large.
The simulations are performed using the stochastic velocity Verlet algorithm introduced in Ref.~\cite{Bus07}, using a time step of $1.25 \cdot 10^{-4}$ and a total time $\tau = 10^3$, after initial equilibration for $\tau_\text{eq} = 10^2$ and average over $100$ repetitions.
We consider two different values for the cutoff length of the interaction potential, $r_c = 4$ (both repulsive and attractive interactions) and $r_c = 2$ (only repulsive interactions), and vary the number of total $N$ and driven $K$ particles between $1$ and $20$ in each case.
The entropy production is computed using \eqref{entropy-force-underdamped}; the results are shown in Fig.~\ref{fig-interacting-entropy}.
We see that for a single particle, the variational expression using one-particle statistics exactly reproduces the entropy production rate, as expected.
Interestingly, the same is observed for the interacting case when all particles are driven; since all particles move coherently with the same velocity, the interactions do not reduce the current.
Conversely, when only some particles are driven, the presence of the passive particles hinders their motion and the current is reduced compared to independent particles.
As a consequence, the entropy production is also reduced and the inequality in \eqref{dem1-upper-bound} becomes strict.
This effect is considerably more pronounced in the presence of attractive interactions, which cause the driven particle to transiently form pairs with passive particles, reducing the mobility of the compound particle and therefore its ability to move in response to the driving force.
In the extreme case of one driven particle in the presence of 19 passive particles, the entropy production is reduced to around half the single-particle bound, compared to a reduction of around 20\% in the purely repulsive case.
As illustrated in Fig.~\ref{fig-interacting-potential}, this corresponds to a situation of almost permanent interaction between the particles.
In conclusion, we find that the simple bound \eqref{dem1-upper-bound} can give a useful estimate of the entropy production even in cases with strong interactions and considerable inertial effects.

\subsection{B{\"u}ttiker-Landauer ratchet}
\begin{figure}
\includegraphics[width=0.47\textwidth]{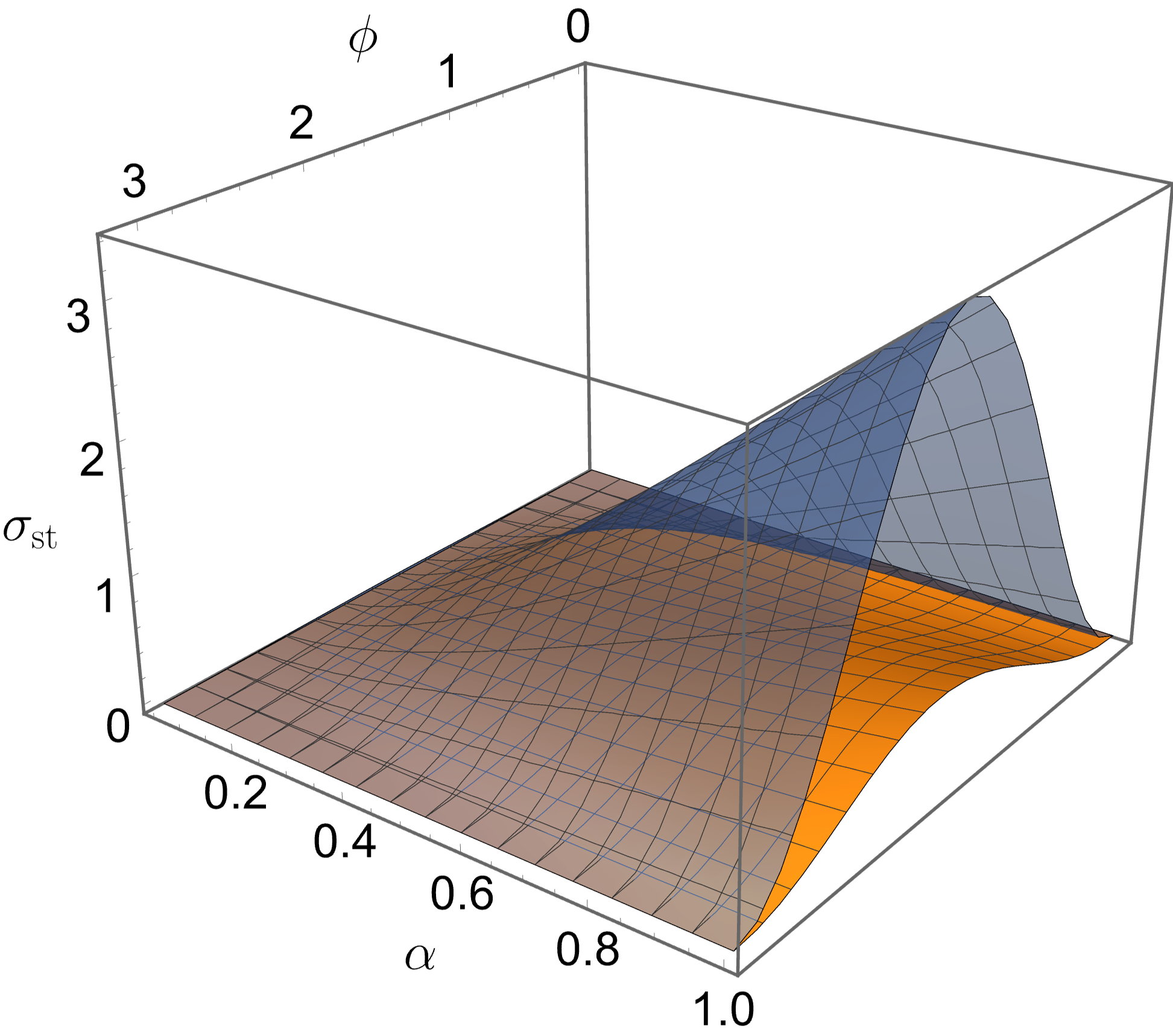}
\caption{The entropy production rate for the B{\"u}ttiker-Landauer with $U_0 = 1$ and $T_0 = 1$, as a function of the magnitude of the temperature variation $\alpha$ and the phase angle $\phi$. The orange surface is the exact entropy production rate \eqref{dem2-entropy}; the blue transparent surface is the upper bound \eqref{entropy-bound-temp-gradient-simple}.} \label{fig-bl-ratchet}
\end{figure}
A simple model that exploits a non-equilibrium steady state in a spatially varying temperature profile to generate motion is the ratchet model introduced by B{\"u}ttiker \cite{Bue87} and Landauer \cite{Lan88}.
Here, an overdamped Brownian particle moves in a one-dimensional periodic potential $U(x+L) = U(x)$ with a likewise periodic spatial temperature profile $T(x+L) = T(x)$.
If the spatial dependence of $U(x)$ and $T(x)$ is chosen appropriately, then this system exhibits a spatially periodic steady state with a non-zero drift velocity $v_\text{st}$.
Due to the one-dimensional nature of this system, all involved quantities can be calculated explicitly, allowing us to compare the bound \eqref{entropy-bound-temp-gradient-simple} to the exact value of the entropy production rate.
As shown in Ref.~\cite{Dec17}, we obtain for the entropy production rate
\begin{align}
&\sigma_\text{st} = \mu \big(1 - e^{\psi(L)}\big)^2 \frac{\int_0^L dx \ e^{\psi(x)} \big[ \int_{x}^{x+L} dy \ e^{\psi(y)} \big]^{-1}}{\int_0^L dx \ e^{\psi(x)} \int_{x-L}^{x} dy \ e^{-\psi(y)}/T(y)}, \nn
&\text{with} \qquad \psi(x) = \int_0^x dy \ \frac{U'(y)}{T(y)} \label{dem2-entropy} . 
\end{align}
As a concrete example, we consider the potential and temperature profile
\begin{align}
U(x) &= U_0 \sin( k x ) \label{dem2-potential} \\
T(x) &= T_0 \bigg( 1 + \frac{\alpha}{2} \big( \sin( k x + \phi ) + 1 \big) \bigg) \n .
\end{align}
Here $k = 2\pi/L$, $U_0$ and $T_0 > 0$ determine the magnitude of the potential and the overall temperature, respectively, $\alpha \geq 0$ determines the magnitude of the spatial temperature variation and $0 \leq \phi < 2 \pi$ is the phase-difference between the potential and temperature profile.
For $\alpha = 0$, the temperature is constant and equal to $T_0$, while for $\alpha = 1$, the temperature varies between $T_0$ and $2 T_0$.
As show in Fig.~\ref{fig-bl-ratchet}, the upper bound correctly captures the qualitative behavior of the entropy production rate: It vanishes for a spatially constant temperature ($\alpha = 0$) and at $\phi = 0$ or $\phi = \pi$, where the potential and the temperature are linearly related (see the discussion in Section \ref{sec-temp-gradient}).
By contrast, both the entropy production and the upper bound are maximal for $\phi = \pi/2$ and increase as the spatial dependence of the temperature becomes more pronounced.
Quantitatively, the actual value of the entropy production rate is around $20 \%$ of the upper bound throughout the range of parameters.
However, we stress that, in contrast to the explicit expression \eqref{dem2-entropy}, computing which requires evaluating the triple integrals in the numerator and denominator, the bound \eqref{entropy-bound-temp-gradient-simple} is easily evaluated from the known expressions \eqref{dem2-potential}.
Moreover, the different reasons for the vanishing entropy production as $\alpha \rightarrow 0$ and as $\phi \rightarrow 0,\pi$ are clearly apparent in the structure of the bound; in the former case, the magnitude $\mathcal{T}$ of the temperature gradient vanishes, while in the latter case $\Delta \mathcal{R}$ vanishes as the relation between temperature and potential becomes linear.

\subsection{Underdamped motion in a periodic potential}
\begin{figure}
\includegraphics[width=0.47\textwidth]{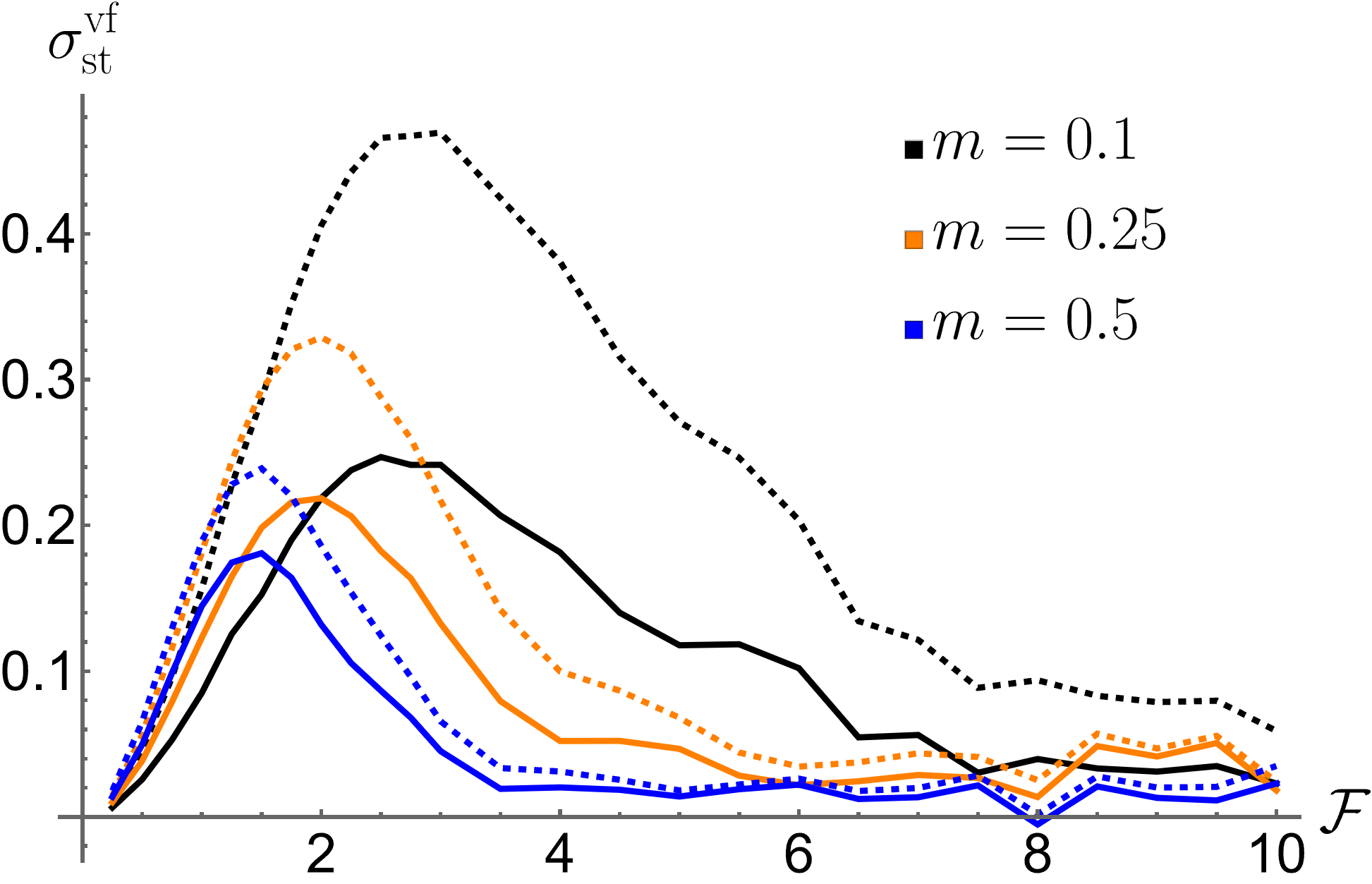}\\
\includegraphics[width=0.47\textwidth]{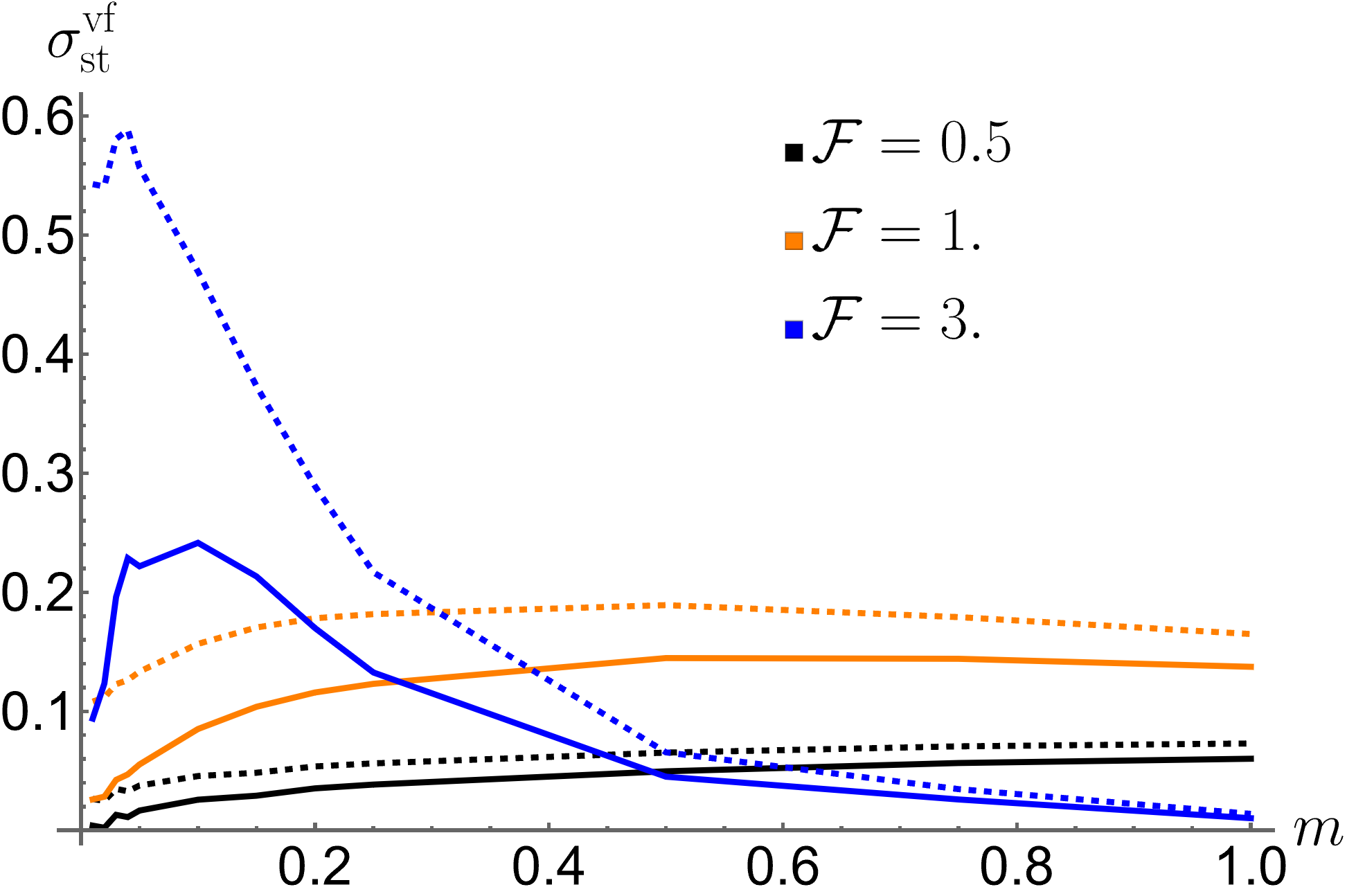}
\caption{The velocity-fluctuation contribution to the entropy production \eqref{dem3-vf-entropy} for an underdamped particle in a tilted periodic potential $U(x) = U_0 \sin(2 \pi x/L)$. The parameters are $\gamma = 1$, $T = 0.5$ and $U_0 = 1$; simulations are performed using $10^4$ trajectories up to $\tau = 10^3 m$, with a time step size of $10^{-4} m$. The solid lines are obtained using \eqref{dem3-vf-entropy} with the drift velocity and position density obtained from the numerical simulations; the dashed lines are the upper bound \eqref{dem3-bound2}.
The top panel shows the results as a function of the driving force $\mathcal{F}$ for three different values of the mass $m$, the bottom panel as a function of the mass $m$ for three different values of the driving force $\mathcal{F}$.} \label{fig-vf-entropy}
\end{figure}
In Section \ref{sec-underdamped}, we found that the upper bounds on the entropy production rate apply to both over- and underdamped dynamics.
In particular, \eqref{velocity-entropy-bound} suggests that the difference between the upper bound and the actual value of the entropy production may be used to estimate the contribution of non-thermal velocity fluctuations to the dissipation.
We study a particle with mass $m$ and friction coefficient $\gamma$ moving in a one-dimensional periodic potential $U(x+L) = U(x)$ and driven by a constant bias force $\mathcal{F}$, so that the Langevin equation \eqref{langevin-underdamped} reads
\begin{align}
\dot{x}(t) &= v(t) \label{dem3-langevin} \\
m \dot{v}(t) &= - U'(x(t)) + \mathcal{F} - \gamma v(t) + \sqrt{2 \gamma T} \xi(t) \n .
\end{align}
We focus on the periodic steady state $p_\text{st}(x,v) = p_\text{st}(x+L,v)$ of the system.
In contrast to the overdamped case, no closed-form expressions for the steady-state density, drift velocity or entropy production rate are available, so we have to rely on numerical simulations of \eqref{dem3-langevin}.
From the simulations, we can measure the drift velocity $v_\text{st}$, which allows us to determine the total entropy production rate $\sigma_\text{st} = v_\text{st} \mathcal{F}/T$.
Next, we note that the optimization problem in \eqref{entropy-variational-under} can be solved explicitly in one dimension.
For a constant driving force, the Euler-Lagrange equation reads,
\begin{align}
\partial_x \Big( p_\text{st}(x) \big( \mathcal{F} + \partial_x V(x) \big) \Big) = 0,
\end{align}
with general solution
\begin{align}
V(x) = c_1 + c_2 \int_0^x dy \ \frac{1}{p_\text{st}(y)} - F x .
\end{align}
The additive constant $c_1$ does not enter \eqref{entropy-variational-under}, while $c_2$ can be obtained by imposing periodic boundary conditions on $V(x)$.
The resulting expression for the upper bound is then
\begin{align}
\hat{\sigma}_\text{st} = \frac{(\mathcal{F} L )^2}{\gamma T} \frac{1}{\int_0^L dx \ \frac{1}{p_\text{st}(x)}} .
\end{align}
We introduce the quantity
\begin{align}
\chi = \frac{L^2}{\int_0^L dx \ \frac{1}{p_\text{st}(x)}} \leq 1 ,
\end{align}
where the inequality follows from the Jensen inequality and equality is attained for a uniform probability density $p_\text{st}(x) = 1/L$.
In terms of this, we have
\begin{align}
\hat{\sigma}_\text{st} = \frac{\mathcal{F}^2}{\gamma T} \chi \leq \frac{\mathcal{F}^2}{\gamma T} \label{dem3-bound},
\end{align}
where the latter expression is obtained by choosing $V(x) = 0$ instead of the actual minimizer.
We can also determine the local mean velocity $\nu_\text{st}(x)$ in terms of $p_\text{st}(x)$.
In one dimension, the solution to the continuity equation \eqref{continuity-underdamped} is
\begin{align}
\nu_\text{st}(x) = \frac{v_\text{st}}{L p_\text{st}(x)},
\end{align}
since the product of $p_\text{st}(x)$ and $\nu_\text{st}(x)$ has to be constant, and its integral over one period equal to $v_\text{st}$.
We then obtain for the local mean entropy production rate defined in \eqref{entropy-decomposition-under},
\begin{align}
\sigma_\text{st}^\text{lm} = \frac{\gamma}{T} \int_0^L dx \ \nu_\text{st}(x)^2 p_\text{st}(x) = \frac{\gamma v_\text{st}^2}{T \chi} .
\end{align}
From \eqref{entropy-decomposition-under}, we therefore obtain the relation
\begin{align}
v_\text{st} \mathcal{F} = \frac{\gamma}{\chi} v_\text{st}^2 + T \sigma_\text{st}^\text{vf} .
\end{align}
In the overdamped case, the last term vanishes, and $v_\text{st} = \chi \mathcal{F}/\gamma$, i.~e., the effective mobility $\mu_\text{eff} = v_\text{st}/\mathcal{F}$ is reduced by a factor $\chi$ compared to the free-space mobility $1/\gamma$.
In the underdamped case, by contrast, we have
\begin{align}
\sigma_\text{st}^\text{vf} = \frac{1}{T} \bigg( v_\text{st} \mathcal{F} - \frac{\gamma}{\chi} v_\text{st}^2 \bigg) \label{dem3-vf-entropy},
\end{align}
which allows us to calculate the contribution of the non-thermal velocity fluctuations to the entropy production, using the measured values of $v_\text{st}$ and $p_\text{st}(x)$.
This also implies that $v_\text{st} \leq \chi \mathcal{F}/\gamma$, i.~e., the drift velocity in the underdamped case is always less than what we would expect from an overdamped system with the same parameters and position density.
On the other hand, evaluating the right-hand side of \eqref{velocity-entropy-bound} using the result \eqref{dem3-bound} for $\hat{\sigma}_\text{st}$, we find precisely the same result for $\sigma_\text{st}^\text{vf}$, so that the upper bound turns into an equality in the one-dimensional case.
This suggests that \eqref{velocity-entropy-bound} can provide a quantitatively useful estimate of the velocity-fluctuation contribution to the entropy production, even without measuring the velocity fluctuations directly.
Evaluating the factor $\chi$ requires determining the position density from the simulation data; \eqref{dem3-bound} also provides an upper bound on the velocity-fluctuation contribution in terms of more easily accessible system parameters and the drift velocity,
\begin{align}
\sigma_\text{st}^\text{vf} \leq \frac{v_\text{st}}{T} \big(  \mathcal{F}  - \gamma v_\text{st} \big) . \label{dem3-bound2}
\end{align}
The values of $\sigma_\text{st}^\text{vf}$ as well as this upper bound are shown in Fig.~\ref{fig-vf-entropy}.
We see that at both small and large driving force, the entropy production due to non-thermal velocity fluctuations vanishes.
For weak driving, the system is close to equilibrium, and thus the velocity distribution becomes thermal.
For strong driving, on the other hand, the system behaves like biased free diffusion, which also exhibits thermal velocity fluctuations relative to the drift velocity.
At intermediate driving, the velocity-fluctuation entropy production exhibits a peak.
Both the driving force corresponding to the peak, as well as the peak value increase with decreasing mass.
The first effect can be understood by noting that, at larger mass, the inertia of the particle leads to so-called running solutions, where the particle essentially continues sliding down the potential for a significant amount of time before becoming trapped in one of the potential wells again.
Thus, the force that is required to approach the free diffusion limit decreases with increasing mass.
The increasing peak value of $\sigma_\text{st}^\text{vf}$ with decreasing mass, on the other hand, is at first glance counterintuitive, since we expect the velocity distribution to become thermal in the overdamped limit.
However, the validity of the overdamped limit requires that the thermalization of the velocity distribution should be smaller than all other timescales associated with the particle's motion.
For increasing force, the time required for the particle to traverse the potential, and, therefore, the characteristic timescale for a change in the force, also decrease, and we have to consider ever smaller masses for the overdamped limit to be valid as we drive the system further from equilibrium.
This is confirmed when considering $\sigma_\text{st}^\text{vf}$ as a function of the mass (bottom panel in Fig.~\ref{fig-vf-entropy}).
At a fixed value of the driving force, the velocity fluctuations do eventually become thermal as we decrease the mass.
However, the eventual decay of $\sigma_\text{st}^\text{vf}$ with decreasing mass occurs at smaller masses for larger forces, leading to a significant deviations from thermal fluctuations even at relatively small mass.
Finally, we note that the upper bound \eqref{dem3-bound2} on the velocity-fluctuation entropy production reproduces this qualitative behavior very well, allowing us to infer the degree of non-thermal velocity fluctuations using only the parameters of the system and the observed steady-state velocity.


\section{Discussion}

Let us first remark on two recent, related results.
First, in Ref.~\cite{Sal22}, an upper bound on the entropy production in terms of the entropy flow has been obtained.
For the class of systems considered here (steady-state systems in contact with thermal environments), however, the entropy flow is already equal to the entropy production, and so the latter bound is only meaningful in more general situations, where the environments are not necessarily thermal.
Second, Ref.~\cite{Nis23} discusses an upper bound on the entropy production in jump processes on a discrete state space, which is expressed in terms of the maximum thermodynamic force in the system.
This result, while obtained in a different physical setting, is closely related to the bounds derived in the present work.
Indeed, from \eqref{entropy-bound-nonconservative-simple} it is straightforward to see that the maximal magnitude of the nonconservative force also yields an upper bound.
Using the variational characterization of entropy production for jump processes derived in Ref.~\cite{Yos23}, it should be possible to obtain more refined versions of the bound in Ref.~\cite{Nis23}.

A fundamental consequence of the upper bound in terms of the one-particle density \eqref{entropy-bound-one-particle} is that the presence of additional, passive degrees of freedom in the system generally leads to a reduction in the dissipation.
As discussed in Section \ref{sec-thermal}, this statement is strictly true only if the interactions do not substantially modify the one-particle density. 
However, it still appears to be in contradiction to the established notion of hidden entropy production \cite{Kaw13,Chu15,Wan16,Nak18}, where neglecting degrees of freedom through coarse-graining causes one to underestimate the dissipation in the system.
This apparent contradiction is resolved by noting that the one-particle bound still includes all the information about the non-conservative force as the source of the dissipation. 
Therefore, \eqref{entropy-bound-one-particle} does not correspond to a coarse-grained description of the system, even though less information (one-particle density vs. many-particle density) is required to compute it compared to the exact entropy production.

In experimental settings, an upper bound on the entropy production rate could be combined with known lower bounds to narrow the range of possible values of the entropy production.
Provided that the thermodynamic forces driving the system out of equilibrium are known, the bounds that we derived here can be computed using various degrees of experimentally obtained information.
If the probability density of the full system is known, then the variational expression \eqref{entropy-variational-nonconservative} reproduces the exact entropy production rate.
On the other hand, if only one-particle statistics are available, we can use \eqref{entropy-bound-one-particle} to derive an upper bound that, as the example in Section \ref{sec-demonstration-interacting} shows, can be relatively tight even in strongly interacting systems.

The upper bounds derived here are also particularly suited for theoretical studies of non-equilibrium phenomena, where we often start from a particular model and want to investigate its behavior and entropy production.
In this case, it can be useful to know the amount of entropy production that the model can exhibit in principle, without solving the entire dynamics.
An upper bound then tells us how far from equilibrium it can potentially be for certain parameters.
This knowledge could then be combined with existing lower bounds to predict the maximal magnitude of non-equilibrium phenomena.
For example, the thermodynamic uncertainty relation \cite{Bar15,Gin16} states that the entropy production bounds the precision of stochastic currents.
Thus, an upper bound on the entropy production also implies a maximal precision of any current in the system, without having to compute the entropy production explicitly.
One interesting detail in this context is that, while it is known that the thermodynamic uncertainty relation can be violated in underdamped systems \cite{Pie22,Dec22b}, the upper bounds on the entropy production hold for both over- and underdamped systems.

Finally, we remark on some possible extensions of the present results.
As mentioned in Section \ref{sec-setup}, most bounds obtained here also apply to the housekeeping entropy production in the Maes-Neto{\v{c}}n{\`y} decomposition \cite{Mae14} for time-dependent driven systems.
It would be interesting to see whether similar upper bounds can also be obtained for the excess part, for example using its known relation with Wasserstein distance \cite{Nak21}, yielding an upper bound on the total entropy production for time-dependent systems.
Extending on the results of Section \ref{sec-multiple-baths}, it may be possible to derive more useful bounds for heat conduction systems, taking into account the structure of the interactions.
Such bounds could be useful to estimate the heat flow, and thus thermal conductance, from the model parameters.
Finally, as remarked in Section \ref{sec-underdamped}, deriving an upper bound for the entropy production in underdamped systems with temperature gradients remains an open yet interesting problem, as such a bound could also provide an estimate on the hidden entropy production that is neglected in the overdamped description.

\begin{acknowledgments}
A.~D.~is supported by JSPS KAKENHI (Grant No. 19H05795, and 22K13974).
\end{acknowledgments}

%

\end{document}